\documentclass[aps,prl,groupedaddress,superscriptaddress,twocolumn,showpacs,longbibliography]{revtex4-1}

\usepackage{graphicx}
\usepackage{dcolumn}
\usepackage{bm}
\usepackage[T1]{fontenc}
\usepackage[french]{babel}
\usepackage{epstopdf}
\usepackage{amssymb}
\usepackage{color}
\usepackage{amsmath}
\usepackage{subfigure}
\usepackage{natbib}

\begin{document}

\preprint{APS/123-QED}

\title{Random strain fluctuations as dominant disorder source for high-quality on-substrate graphene devices}

\author{Nuno J. G. Couto}
\author{Davide Costanzo}
\affiliation{ D\'{e}partement de Physique de la Mati\`{e}re
Condens\'{e}e (DPMC) and Group of Applied Physics (GAP), University
of Geneva, 24 Quai Ernest-Ansermet 1211 Gen\`{e}ve 4, Switzerland}

\author{Stephan Engels}
\affiliation{JARA-FIT and 2nd Institute of Physics, RWTH Aachen
University, 52074 Aachen, Germany, and Peter Gr\"{u}nberg Institute
(PGI-9), Forschungszentrum Jülich, 52425 Jülich, Germany}

\author{Dong-Keun Ki}
\affiliation{ D\'{e}partement de Physique de la Mati\`{e}re
Condens\'{e}e (DPMC) and Group of Applied Physics (GAP), University
of Geneva, 24 Quai Ernest-Ansermet 1211 Gen\`{e}ve 4, Switzerland}

\author{Kenji Watanabe}
\author{Takashi Taniguchi}
\affiliation{National Institute for Materials Science, 1-1 Namiki, Tsukuba 305-0044, Japan}

\author{Christoph Stampfer}
\affiliation{JARA-FIT and 2nd Institute of Physics, RWTH Aachen
University, 52074 Aachen, Germany, and Peter Grünberg Institute
(PGI-9), Forschungszentrum J\"{u}lich, 52425 J\"{u}lich, Germany}

\author{Francisco Guinea}
\affiliation{Instituto de Ciencia de Materiales de Madrid, CSIC, Cantoblanco, E-28015 Madrid, Spain}

\author{Alberto F. Morpurgo}
\affiliation{ D\'{e}partement de Physique de la Mati\`{e}re
Condens\'{e}e (DPMC) and Group of Applied Physics (GAP), University
of Geneva, 24 Quai Ernest-Ansermet 1211 Gen\`{e}ve 4, Switzerland}


\begin{abstract}
We have performed systematic investigations of transport through
graphene on hexagonal boron nitride (hBN) substrates, together with
confocal Raman measurements and a targeted theoretical analysis, to
identify the dominant source of disorder in this system.
Low-temperature transport measurements on many devices reveal a
clear correlation between the carrier mobility $\mu$ and the width
$n^*$ of the resistance peak around charge neutrality, demonstrating
that charge scattering and density inhomogeneities originate from
the same microscopic mechanism. The study of weak-localization
unambiguously shows that this mechanism is associated to a
long-ranged disorder potential, and provides clear indications that
random pseudo-magnetic fields due to strain are the dominant
scattering source. Spatially resolved Raman spectroscopy
measurements confirm the role of local strain fluctuations, since
the line-width of the Raman 2D-peak --containing information of
local strain fluctuations present in graphene-- correlates with the
value of maximum observed mobility. The importance of strain is
corroborated by a theoretical analysis of the relation between $\mu$
and $n^*$ that shows how local strain fluctuations reproduce the
experimental data at a quantitative level, with $n^*$ being
determined by the scalar deformation potential and $\mu$ by the
random pseudo-magnetic field (consistently with the conclusion drawn
from the analysis of weak-localization). Throughout our study, we
compare the behavior of devices on hBN substrates to that of devices
on SiO$_2$ and SrTiO$_3$, and find that all conclusions drawn for
the case of hBN are compatible with the observations made on these
other  materials. These observations  suggest that random strain
fluctuations are the dominant source of disorder for high-quality
graphene on many different substrates, and not only on hexagonal
boron nitride.
\end{abstract}

\maketitle
\section{Introduction}
Hexagonal boron nitride (hBN) substrates enable the fabrication of
graphene
devices~\cite{Dean2010,Zomer2011_TranshBN,Mayorov2011_micron,Wang2013},
exhibiting extremely high carrier mobility values, and leading to
the observation of new, interesting physical
phenomena~\cite{dean2011_FQHE,Young2012_svqHF,Ponomarenko2013_Cloning,Dean2013_hof,Hunt2013_hof}.
The precise microscopic reason for the quality of these devices,
however, has not yet been established, nor is it understood what is
the dominant microscopic physical mechanism responsible for the
remnant disorder. Here, we perform a systematic study of a large
number of such devices, and provide considerable evidence --both
experimentally and theoretically-- that random local strain
fluctuations in the graphene lattice are the dominant microscopic
source of disorder.

Many different techniques are currently used for the production of
graphene devices, and the dominant source of disorder depends on the
specific type of device considered. We confine our attention to high
quality devices, based on graphene monolayers exfoliated from
natural graphite and transferred to be in direct contact with a
substrate material, not exposed to damaging agents (such as electron
or ion beams, ultra violet radiation, or aggressive chemical
environments). Even so, many different physical mechanisms --such as
charged impurities at the substrate surface, adsorbates acting as
resonant scatterers, structural defects such as vacancies, strain
fluctuations, and more-- have been considered as possible sources of
disorder~\cite{CastroNeto2009}. Conducting targeted experiments to
identify the dominant source in any given individual device is
virtually impossible, and information can only be extracted by
analyzing the statistical behavior of many devices realized under
controlled conditions. Experiments have been performed to
intentionally introduce one specific type of disorder in graphene
(e.g., charged impurities, by depositing an increasingly large
number of potassium atoms on a graphene layer~\cite{Chen2008_cimp},
or vacancies, by bombarding graphene with an increasingly large dose
of heavy ions~\cite{Chen2009_vacancy}) while monitoring the
resulting variations in the electronic properties. This work has
allowed testing  specific predictions of theories describing
disorder of different nature, but has not enabled the determination
of the physical mechanism causing the disorder initially present in
the devices. Considerable research has been devoted to analyze the
dependence of the conductivity of graphene ($\sigma$) on carrier
density ($n$), without, however, solving the existing controversies,
mainly because the measured $\sigma(n)$ curves are consistent with
the functional dependence obtained from models describing different
sources of disorder.  Despite the work of many different research
groups, there is not even established consensus for the most common
devices on SiO$_2$, as to whether the dominant disorder potential is
short or long-ranged (i.e., whether it has  a range comparable to
the lattice spacing or much
longer)~\cite{Adam2007_Pnas,Jang2008,Ponomarenko2009_highK,Peres2010_review,Monteverde2010_prl,
DasSarma2011_review,Couto2011,Aetal11}.

Our work exploits a combination of different experimental
techniques, together with the statistical analysis of a large number
of devices on hBN substrates, looking at both the carrier mobility
$\mu$ and the width of the resistance peak around charge neutrality
$n^*$. While the best graphene-on-hBN devices exhibit impressively
high mobility values, more modest values are also commonly found, so
that the resulting broad range of electrical characteristics allows
the identification of correlations between different quantities. We
find an unambiguous correlation between the carrier mobility $\mu$
and the width of the resistance peak around charge neutrality $n^*$
--with $\mu \propto(n^*)^{-1}$-- extending  over nearly two orders
of magnitude, which demonstrates that the physical mechanism
limiting the mobility is the same one causing charge inhomogeneity.
To identify this mechanism, we perform weak-localization
measurements to extract several characteristic scattering times,
such as the inter-valley scattering time $\tau_{iv}$ and the time
$\tau_{*}$ associated to the breaking of the effective,
single-valley time reversal symmetry. For all charge carrier
densities, $\tau_{iv}\gg \tau$, the elastic scattering time
extracted from the carrier mobility. This finding directly
establishes that the mobility is limited by intra-valley scattering
caused by long-ranged potentials, confining the possible microscopic
mechanisms to charged impurities and random strain fluctuations in
the graphene lattice.

Two independent observations indicate that local strain fluctuations
dominate. First, weak-localization measurements show that $\tau_{*}$
and $\tau$ nearly coincide, a finding that is readily explained if
pseudo-magnetic fields due to local strain are the dominant source
of elastic scattering, but that cannot be explained by the charged
impurity mechanism. Second, we directly probe local strain
fluctuations with confocal Raman experiments~\cite{neu14}, and show
experimentally that larger strain fluctuations limit the maximum
mobility that can be observed in transport measurements. Based on
this evidence, we analyze theoretically the linear relation between
$1/\mu$ and $n^*$ --which had been previously observed in devices
exposed to potassium atoms, and taken to be an indication of charge
impurity scattering-- and show that such a relation can be explained
quantitatively invoking random strain fluctuations only. According
to this same analysis, it is the random pseudo-magnetic field
originating from strain fluctuations, and not the deformation
potential, that gives the dominant contribution to the scattering of
charge carrier, in agreement with the conclusion drawn from the
analysis of weak-localization. Whereas most of our work has focused
on graphene-on-hBN devices, we also have looked at devices on
SiO$_2$ and SrTiO$_3$ substrates and found that the observations
made on these devices are fully compatible with the conclusions
drawn for hBN, which points to the relevance of random strain
fluctuations under rather broad experimental conditions for
high-quality graphene devices on different substrates.

\section{Extracting $\mu$ and $n^*$ for graphene devices on hBN}
The fabrication of graphene-on-hBN devices relies on a technique
described in the literature~\cite{Dean2010}. We exfoliate hBN
crystals onto a heavily doped, oxidized Si wafer. Graphene flakes
extracted from natural graphite are transferred onto a hBN crystal,
following the procedure of Ref.~\cite{Dean2010}. Metallic contacts
(Ti/Au, 10/75 nm) are defined by electron-beam lithography,
evaporation and lift-off (see Fig. 1(a)). We find that "bubbles" and
"folds" form when transferring graphene on hBN (as in
Ref.~\cite{Mayorov2011_micron,Taychatanapat2013_Bfocus,Zomer2011_TranshBN})
: achieving high-$\mu$ requires etching Hall bar devices in parts of
the flakes where no such defects are present (regions with "bubbles"
exhibit lower $\mu$, comparable to SiO$_2$ devices). After an
electrical characterization at 4 K, we perform different
low-temperature thermal annealing steps (at up to 150-250 $^\circ$C,
in an environment of H$_2$/Ar at 100/200 sccm) and check each time
the low-temperature transport characteristics. We find that the
initial annealing step always results in a mobility increase (a
factor of 2 in the very best cases), whereas subsequent annealing
lead to a decrease in $\mu$, eventually to values similar to those
obtained on SiO$_2$~\cite{Chen2008}.

We analyzed approximately 15 distinct Hall-bar devices. Mobility
values (at 4.2 K) between 30.000 cm$^{2}$/Vs and 80.000 cm$^{2}$/Vs
at a carrier density of a few $10^{11}$ cm$^{-2}$ were found
regularly. Integer quantum Hall (QH) plateaus with  $\sigma_{Hall} =
4(1/2 + N)e^2/h$ ($N$ integer) are fully developed starting from
$B=1$ T, and broken symmetry QH states with Hall conductivity
$\sigma_{Hall} = \pm 1e^2/h$  appear from $B=8$ T. Full degeneracy
lifting of the $ N = 0$ and $N=\pm 1$ Landau levels is  observed
below 15 T (Fig. 1(e)). In devices where the lattices of graphene
and hBN were intentionally aligned, we  observe the effect of a
superlattice potential, with the appearance of satellite Dirac peaks
in the measured $R(V_g)$ curve (Fig.
1(c))~\cite{Ponomarenko2013_Cloning,Dean2013_hof,Hunt2013_hof}.
These results indicate that our devices have quality comparable to
those fabricated using a similar procedure, reported in the
literature.

\begin{figure}[t]
\begin{center}
\includegraphics[width=1\linewidth]{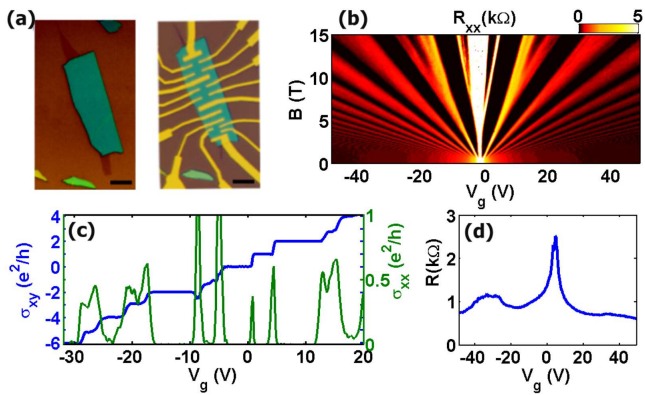}
\caption{(Color online) (a) Optical microscope image of a monolayer
graphene flake on top of a 30 nm thick hBN crystal before (left) and
after (right) depositing metal contacts (the scale bars are 5
$\mu$m). (b) Longitudinal resistance $R_{xx}$ as a function of $V_g$
and $B$ showing quantum Hall states originating from the lifting of
the single-particle degeneracy already at $B\simeq 8$~T. (c) Hall
(blue) and longitudinal (green) conductivity as a function of $V_g$
measured at $B=15$~T, showing full degeneracy lifting of Landau
level $N = 0,1$. (d) Resistance of a graphene device whose edge was
aligned to that of the hBN substrate, showing the emergence of
satellite Dirac peaks (well developed for negative $V_g$ and less
pronounced for positive $V_g$). All measurements have been taken at
$T=250$ mK.}
\end{center}
\label{FIG.1}
\end{figure}

To evaluate  the quality of our graphene-on-hBN devices we focus on
the low-$T$ mobility $\mu$ and on the width $n^*$ of the minimum in
the conductivity $\sigma$-vs-$V_g$ curve. The mobility $\mu$
measures the elastic scattering time $\tau$ responsible for momentum
relaxation, whereas $n^*$ quantifies the potential fluctuations
experienced by electrons in
graphene~\cite{Martin2008,CastroNeto2009}. Since these potential
fluctuations are not a priori the dominant source of elastic
scattering, there is no reason to assume that $\mu$ and $n^*$ are
related. Experimentally, the carrier mobility is obtained from
$\mu=\sigma/ne$ (see Fig. 2(a)), with the density of charge carriers
$n$ obtained through the Hall resistance. To extract $n^*$ we plot
$\log(\sigma)$ as a function of $\log (n)$, and determine at which
$n$ the constant value of $\log(\sigma)$ measured at low density
crosses the value of $\log(\sigma)$ extrapolated (linearly) from
high density (as shown in Fig. 2(b)). The mobility is estimated for
$n>n^*$.

\begin{figure}[t]
\begin{center}
\includegraphics[width=1\linewidth]{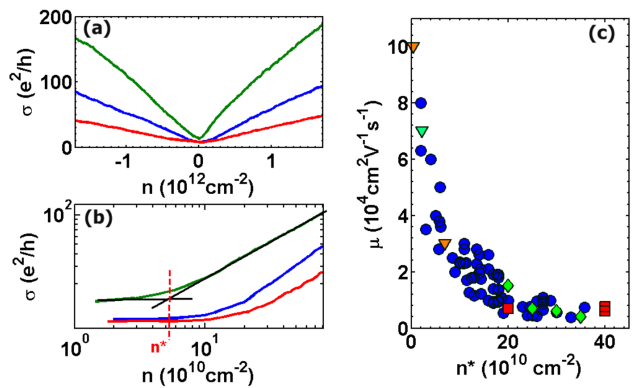}
\caption{(Color online) Conductivity $\sigma$ of a graphene
monolayer on hBN as a function of carrier density $n$ in linear (a)
and double-logarithmic (b) scale, measured  after fabrication (blue
line), after a first annealing at 150 $^\circ$C (green line), and
after a second annealing at 250 $^\circ$C (red line). Panel (b) also
illustrates the procedure to extract the value of $n^*$. (c) The
blue full circles represent the low-temperature mobility $\mu$
(plotted versus $n^*$) for all the 15 graphene-on-hBN devices
realized in our laboratory, measured after fabrication or after
annealing. The triangles represent data for graphene-on-hBN
extracted from Ref.~\cite{Dean2010,dean2011_FQHE} (orange triangles)
and from Ref.~\cite{Zomer2011_TranshBN} (green triangle). The green
diamonds and red squares are from devices realized in our laboratory
on SiO$_2$ and SrTiO$_3$ substrates, respectively.}
\end{center}
\label{FIG.2}
\end{figure}

Fig. 2(c) shows $\mu$ as a function of $n^*$ for all devices,
measured either immediately after fabrication, or after a subsequent
annealing step. The presence of a correlation between $\mu$ and
$n^*$ is unambiguous: devices with smaller density fluctuations have
larger mobility. For hBN devices fabricated in our laboratory, this
correlation extends from $\mu$ values of 5.000 cm$^2$/Vs (for
devices after multiple annealing steps, see below) to 80.000
cm$^2$/Vs. Results reported in the
literature~\cite{Dean2010,Zomer2011_TranshBN,dean2011_FQHE}
quantitatively fit the same trend, extending the range  to $\mu =
100.000$ cm$^2$/Vs. Plotting  $1/\mu$-vs-$n^*$ (Fig. 3(a)) shows
that the relation between these two quantities is essentially
linear. To reduce the statistical fluctuations we subdivide the
$n^*$ axis into eight different intervals and plot the inverse
averaged mobility as a function of the average charge density
fluctuations (Fig. 3(b)), which makes the linear scaling of $1/\mu$
with $n^*$  apparent.

We emphasize that neither the occurrence of the relation between
$1/\mu$ and $n^*$, nor its approximate linearity, are obvious a
priori. Indeed, it has been shown that when intentionally creating
carbon vacancies, no such relation is observed, because in that case
vacancies are the dominant mechanism responsible for the suppression
of the carrier mobility, but they are not the dominant mechanism
causing charge inhomogeneity~\cite{Chen2009_vacancy}. Our
observations, therefore, unambiguously establish that scattering of
charge carriers and charge inhomogeneity in devices on hBN are
caused by the same microscopic mechanism. Also the linearity of the
$1/\mu$-vs-$n^*$ relation is not trivial: we have measured several
graphene bilayer devices on hBN and SiO$_2$ and found that a
relation between $1/\mu$ and $n^*$ occurs also in that case, but the
the relation is quadratic and not linear (see Appendix B). These
considerations make clear that a quantitative analysis of the
$1/\mu$-vs-$n^*$ relation can provide important information. Note
that a correlation similar to the one shown in Fig. 3(a,b) has been
reported for graphene covered by ionized potassium atoms, which do
generate disorder consistent with the charged impurities
mechanism~\cite{Chen2008_cimp}. On this basis, one may be tempted to
conclude that charged impurities are also the dominant source of
disorder for graphene on hBN. As we will show below, however, the
$1/\mu$-vs-$n^*$ correlation is also qualitatively and
quantitatively compatible with the effect of random strain
fluctuations in graphene, and discriminating between charged
impurities and strain is the main goal of the remaining part of this
paper. Before coming to that, we notice that, rather surprisingly,
the $1/\mu$-vs-$n^*$ correlation is fulfilled also by devices on
different substrate materials, whose data point --the red and green
dots in Fig. 2(c) represent data obtained from graphene on SiO$_2$
and SrTiO$_3$~\cite{Couto2011}-- fall on the curve defined by the
results obtained for graphene devices on hBN~\footnote{In
Ref.~\cite{Couto2011} we discussed transport through
graphene-on-SrTiO$_3$ in the context of resonant scattering, but we
also pointed out --in Ref.~(21) of that paper-- that the data are
compatible with scattering by ripples, i.e.  with the conclusions of
this present work.}.

\begin{figure}[t]
\begin{center}
\includegraphics[width=1\linewidth]{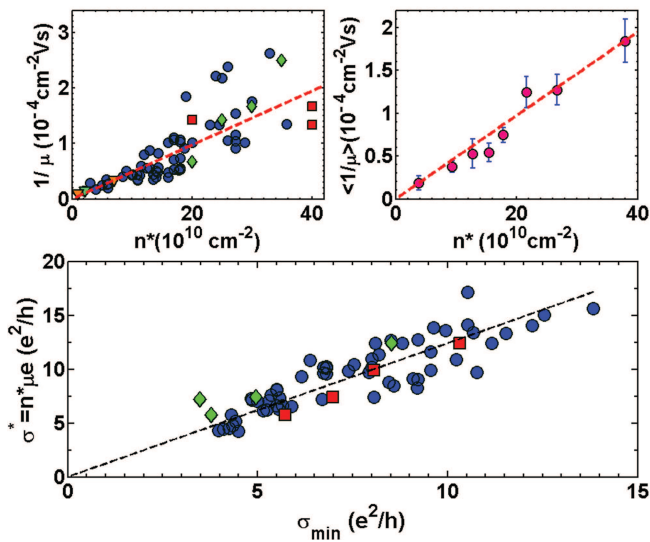}
\caption{(Color online)(a) Same data as those of Fig. 2(c) plotted
as $1/\mu$-vs-$n^*$, showing an overall linear relation. (b) Average
inverse mobility as function of $n^*$ (obtained as indicated in the
main text), showing clearly the linearity of the relation. In (a)
and (b) the dashed lines are a linear fit to the data,
$\frac{1}{\mu} = \frac{h}{e}n^* \times 0.118$. (c) minimum
conductivity $\sigma^*=n^*e\mu$, calculated from the estimated
carrier density fluctuations $n^*$ and mobility $\mu$, and plotted
as a function of measured minimum conductivity. The excellent
overall agreement (the dashed line has slope 1) confirms the
correctness of the procedures used to extract $n^*$ and $\mu$ from
the measurements.}
\end{center}
\label{FIG.3}
\end{figure}

\section{The characteristic scattering times reveal the origin of disorder}
Having established that scattering of charge carriers and carrier
density inhomogeneities are caused by the same microscopic
mechanism, we can gain additional insight by analyzing
weak-localization to extract all the relevant scattering times for
graphene on hBN~\cite{McCann2006,Morpurgo2006,Tikhonenko2008}. Our
first goal is to compare the inter-valley scattering time
$\tau_{iv}$ to the elastic scattering time $\tau$ determined from
the carrier mobility. Either $\tau_{iv} \simeq \tau$, implying that
the mobility is determined by inter-valley scattering processes
(i.e., the dominant source of disorder are short-range potentials),
or $\tau_{iv} \gg \tau$, indicating that $\mu$ is limited by
intra-valley scattering (i.e., long-range disorder potentials
dominate). Surprisingly, this straightforward argument has not been
used systematically in previous work to identify the dominant
disorder, nor has it been suggested in theoretical work (for an
exception, see Ref.~\cite{Guignard2012} dealing with rather low
mobility devices, $\mu \simeq 1.000$ cm$^2$/Vs).

Fig. 4(a) shows the low-field magneto-resistance of a Hall bar
device with $\mu \simeq 60.000$ cm$^2$/Vs, for different values of
$V_g$ around $V_g=8$ V, at $T=250$ mK. A narrow dip in conductivity (width $\simeq
1$ mT or less) is seen around $B=0$~T, originating from weak
localization. Aperiodic conductance fluctuations due to random
interference are also visible, which we suppress by averaging
measurements taken for slightly different $V_g$
values~\cite{FerryBook2009}. "Ensemble-averaged" curves obtained in
this way around three different $V_g$ values are shown in Fig. 4(b).
We have performed similar measurements at several different temperatures,
and analyzed the ensemble-averaged low-field magneto-transport up to $T=10$ K.

To analyze the data, we have followed the same procedure used in
previous studies of the quantum correction to the conductivity done
on graphene on SiO$_2$ substrates~\cite{Tikhonenko2008,Guignard2012}
and on epitaxial graphene on SiC~\cite{Baker2012}. Specifically, the
data are fit to existing theory~\cite{McCann2006}, from which we
extract the inter-valley scattering time $\tau_{iv}$, the phase
coherence time $\tau_{\phi}$, and the time $\tau_{*}$ needed to
break effective single-valley time reversal
symmetry~\cite{Morpurgo2006}, using the equation
\begin{eqnarray}\nonumber
\Delta \sigma(B)& =& \frac{e^2}{\pi h}\left(F\left(\frac{\tau^{-1}_B}{\tau^{-1}_\phi}\right) -F\left(\frac{\tau^{-1}_B}{\tau^{-1}_\phi + 2\tau^{-1}_{iv}}\right)\right. \\
 && \left. - 2F \left(\frac{\tau^{-1}_B}{\tau^{-1}_\phi + \tau^{-1}_{iv} + \tau^{-1}_*}\right)\right),   \label{WLeq}
\end{eqnarray}
Here $F(z)=\ln z+\psi(0.5+z^{-1})$ ,$\psi(x)$ is the digamma,
function, $\tau_B^{-1} = 4eDB/\hbar$ and $D = v_F \tau/2$. In
fitting the magneto-transport curves at different temperatures, we
allow $\tau_\phi$ to vary --since the phase coherence time does
increase with lowering $T$-- and we constrain the other scattering
times to be constant in the range investigated (250 mK and 10 K). We
obtain satisfactory agreement in all cases with a single set of
value for $\tau_{iv}$ and $\tau_{*}$ (the elastic scattering time
$\tau$ --also constant as a function of $T$-- is obtained from the
measurements of the conductivity, and is not a fitting parameter).

Fig. 4(c) shows the hierarchy of the relevant times at $T=250$ mK,
the lowest temperature reached in the experiments, for three
different values of $n$. At this temperature, $\tau_{\phi}$ is much
larger than $\tau_{iv}$, which is why weak-localization is observed
($\tau_{\phi}$ eventually becomes shorter than $\tau_{iv}$ as $T$
reaches 10 K, so that  weak antilocalization becomes visible, in
conformity to theoretical expectations, and as found previously for
graphene on SiO$_2$~\cite{Tikhonenko2008,Guignard2012}). More
importantly throughout the density range investigated $\tau_{iv} \gg
\tau$ by at least one order of magnitude, (and by nearly two at low
$n$). This last observation implies that intra-valley scattering is
the process limiting $\mu$, a result that --in conjunction with
previous measurements on
graphene-on-SiO$_2$~\cite{Tikhonenko2008,Guignard2012}-- holds at
least in the mobility range between 1.000 and 80.000 cm$^2$/Vs. We
conclude that weak-localization measurements unambiguously show that
the dominant source of disorder for exfoliated mono-layer graphene
on hBN (and SiO$_2$) substrates is associated to long-ranged
potentials (motivated by this conclusion, we have also recently
studied weak-localization  on high-quality graphene bilayer devices
on hBN substrates, and in that case as well we have unambiguously
come to the same conclusion, namely that it is intra-valley
scattering processes that are  limiting the carrier
mobility~\cite{engels14}).

The results of the weak-localization measurements also provide a
clear indication as to which of the two sources of long-range
disorder (charged impurities at the substrate
surface~\cite{Adam2007_Pnas,Ando,Nomura2007} and random strain
fluctuations in the graphene lattice~\cite{katsnelson_corrug}) plays
the most relevant role. Specifically, the analysis of
weak-localization shows that $\tau \simeq \tau_*$ within a factor of
2-3, for all carrier density range investigated (Fig. 4(b)), a
finding that is naturally explained by strain. Indeed, strain
generates random pseudo-magnetic fields~\cite{Vozmediano2010} that
not only scatter charge carriers, but also break the effective time
reversal symmetry in a single-valley~\cite{Morpurgo2006,McCann2006}
on approximately the same time scale. If these random
pseudo-magnetic fields are the dominant source of scattering
limiting the mobility, we can immediately understand why $\tau$ and
$\tau_*$ are comparable. On the contrary, for a potential $V$
generated by charged impurities on and in the substrate, $\tau$ is
determined by the Fourier components $V(k)$ with  $k \approx k_F$,
whereas $\tau_*$ is determined by random fluctuations in the
potential difference between the A and B atoms in the individual
unit cells of graphene, i.e. by the Fourier component of $V$ with $k
\simeq 1/a$ (see Appendix C). Since $V$ is a long-range potential,
$V(k_F)\gg V(1/a)$, implying (through Fermi golden rule) that for
charged impurities  $\tau_* \gg \tau$, in disagreement with the
experimental observations. We are not aware of any mechanism other
than strain-induced pseudo-magnetic fields that can explain the
coincidence between $\tau$ and $\tau_*$, which is why the indication
of this finding for the relevance of local strain fluctuations is
rather compelling.

Albeit less directly, the experimentally observed evolution of $\mu$
upon annealing also points to the effect of strain. As discussed
above, repeated annealing at low temperature ($\simeq 200$
$^\circ$C) in an inert atmosphere systematically reduces $\mu$ by
one order of magnitude. These annealing processes have no
significant chemical effect, and therefore are not expected to
change the density of charge at the surface of hBN by one order of
magnitude (as it would be needed to explain the changes in
$\mu$~\cite{Adam2007_Pnas}). On the contrary, they do lead to
visible mechanical deformations, compatible with strain causing a
decrease in mobility. Finally, having $\mu$ limited by
strain-induced pseudo-magnetic fields also explains why the use of
high-$\epsilon$ substrates --such as SrTiO$_3$~\cite{Couto2011}--
does not lead to a very large increase in mobility: a
high-$\epsilon$ substrate can screen scalar potentials, but not the
effect of a pseudo-magnetic field.

\begin{figure}[t]
\begin{center}
\includegraphics[width=1\linewidth]{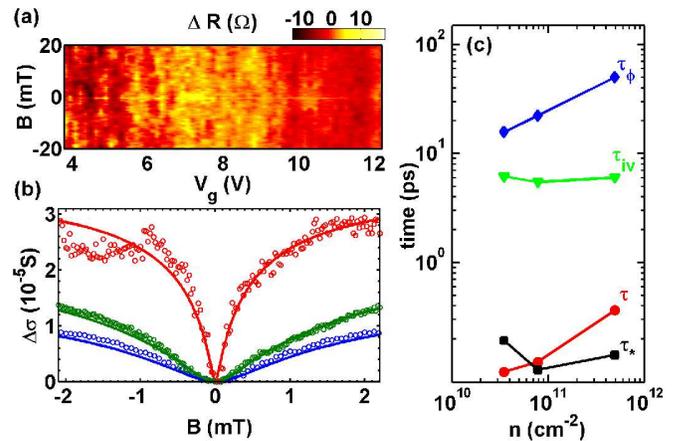}
\caption{(Color online)(a) $B$ and $V_g$ dependence of the
resistivity measured at $T=250$~mK. (b) The circles represent
magneto-conductivity curves $\Delta \sigma(B)$ that have been
ensemble averaged, by averaging traces in a range of gate voltages
around $V_g=-7$ (blue circles), 7 (green circles), and 30 V (red
circles), to suppress sample specific fluctuations. The continuous
lines are fit to the theory of weak-localization in graphene. (c)
Characteristic times extracted at 250 mK for different values of
carrier density, either from the fit of weak localization curves
($\tau_{\phi}$, $\tau_{iv}$, and $\tau_{*}$) or from the
conductivity ($\tau$). The elastic scattering time  $\tau$ is always
at least one order of magnitude smaller than the inter-valley
scattering time $\tau_{iv}$.}
\end{center}
\label{FIG.4}
\end{figure}

\section{Raman mapping for correlating strain fluctuations and
carrier mobility} Additional indications that carrier mobility in
graphene is limited by local strain fluctuations can be obtained by
combining transport measurements with spatially resolved Raman
spectroscopy~\cite{graf07,lee12}. The quantity of interest in this
case is the line-width of the Raman 2D-peak, $\Gamma_{2D}$. In
contrast to the width of the G-peak, $\Gamma_{2D}$ does only very
weakly depend on doping, charge
inhomogeneities~\cite{pisana07,stampfer07,berciaud13}, or magnetic
field~\cite{neu14}. $\Gamma_{2D}$ is also only weakly affected by
global strain and by the different screening properties of the
substrates~\cite{popov13,forster13}, while  it is highly sensitive
to strain inhomogeneities on length-scales smaller the laser-spot
size ($<500$ nm), as recently shown by Neumann and
coworkers~\cite{neu14}. These are precisely the random strain
fluctuations that can contribute to scattering of charge carriers.

\begin{figure}[t]
\includegraphics[width=0.48\textwidth]{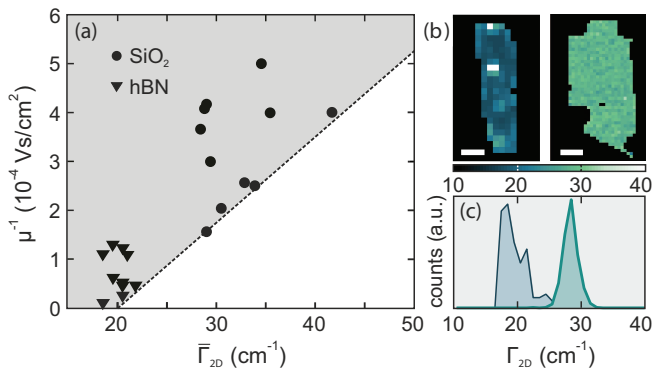}
\caption{(color online) (a) Correlation of the inverse mobility and
the average full width at half maximum of the  Raman 2D-peak,
$\Gamma_{2D}$ for a number of graphene flakes on different
substrates. (b) Raman maps of two graphene flakes resting on two
different substrates (left: hBN and right: SiO$_2$) highlighting the
different values of the spatially resolved $\Gamma_{2D}$ (same color
scale). The white scale bars are 2~$\mu$m. (c) Histograms of
$\Gamma_{2D}$ for the two examples shown in (b). These histograms
are used to extract the data points illustrated in panel (a).}
\label{ramanfig}
\end{figure}

Fig.~\ref{ramanfig}(a) shows the inverse mobility $\mu^{-1}$ versus
the line-width of the 2D-peak for a number of contacted graphene
flakes resting on different substrates. Each of the data points
corresponds to a different sample, on which we performed
low-temperature ($T$=1.8~K) transport measurements to extract the
mobility $\mu$, as well as spatially resolved Raman maps, such as
those of Fig.~\ref{ramanfig}(b) (the color code represents the
line-width $\Gamma_{2D}$ of the local 2D-peak). From these maps we
extract for each flake the distribution of $\Gamma_{2D}$, see
Fig.~\ref{ramanfig}(c), from which we calculate the average width
$\bar{\Gamma}_{2D}$ (this is the quantity plotted on the horizontal
axis of Fig.~\ref{ramanfig}(a)). Values of $\bar{\Gamma}_{2D}$
larger than the intrinsic line-width of the 2D-peak are indicative
of strain fluctuations in the graphene layer~\cite{neu14}, and a
larger $\bar{\Gamma}_{2D}$ corresponds to a larger magnitude of
these random strain fluctuations.

Finding that the data points from all the investigated devices in
Fig.~\ref{ramanfig}(a) lie above the dotted lines means that the
maximum observed value of $\mu$ is smaller in devices for which
$\bar{\Gamma}_{2D}$ is larger, i.e., in devices with larger random
strain fluctuations. This directly indicates that strain is limiting
the carrier mobility. The data show a rather large spread in
mobility values, which originates from the fact that the mobility
can be limited by structural defects --like folds formed in graphene
during the transfer and fabrication process-- which can have only a
small effect on the averaged linewidth $\bar{\Gamma}_{2D}$. Indeed,
the devices used for these combined Raman and transport measurements
were not etched to confine transport through regions in which these
types of structural defects are absent, since etching would have
drastically reduced the area of graphene, making Raman measurements
considerably more complex. As a result, a quasi one-dimensional fold
or ripple cutting across the graphene flake (see e.g. white regions
in Fig.~\ref{ramanfig}(b), left panel) can have a very strong effect
on the mobility value extracted in the device, while --as it affects
only a small part of the total device area-- it has only a small
effect on the averaged linewidth $\bar{\Gamma}_{2D}$. Despite these
experimental limitations, the {\em absence} of data points in the
non-shaded area  indicates that a necessary condition to observe
high carrier mobility values is to have small random strain
fluctuations, and the correlation between maximum mobility and
averaged linewidth $\bar{\Gamma}_{2D}$ is clearly apparent in the
data.

\section{Quantitative explanation of the $1/\mu$-vs-$n^*$ correlation
in terms of strain} Having found a direct correlation between the
strength of the random local strain in graphene and the carrier
mobility --and therefore having confirmed the role of strain
fluctuations as an important source of disorder-- we check, for
consistency, whether the relation between $1/\mu$ and $n^*$  that we
discussed earlier (see Fig. 3(a,b)) can be explained theoretically
in terms of strain fluctuations only. As we mentioned already, such
a relation has been reported experimentally earlier on, in the study
of transport through graphene exposed to an increasingly large
density of potassium atoms, where it was naturally explained in
terms of the effect of  charged impurities (the ionized potassium
atoms)~\cite{Chen2008_cimp}. Here below we show that the relation
between $1/\mu$ and $n^*$ is very naturally reproduced also if
random local strain is the dominant source of disorder. Indeed, at
the quantitative level, the experimental data agree with theoretical
calculations for realistic values (i.e., in the range known from
literature) of the elastic parameters of graphene, which describe
the coupling between strain and electronic properties.

Strain can originate from both in-plane and out-of-plane
deformations (the latter being the so-called ripples), with the
former being probably the most relevant ones, especially on hBN
substrates. The effect of random strain fluctuations on the motion
of electrons in graphene can be described by introducing a scalar
and a vector potential  $V_s$ and $\mathbf{A}$ in the
long-wavelength Dirac Hamiltonian. What is needed to calculate the
effect of strain fluctuations on $\mu$ and $n^*$ are the correlation
functions of these potentials, which can all be obtained directly
from the correlation function of the random strain field (as
described in Appendix A). The scalar and the gauge potential scatter
electrons (with rates $1/\tau_s$ and $1/\tau_g$, respectively) and
limit the mobility $\mu$. The magnitude of the charge fluctuations
$n^*$, on the contrary, is determined by the scalar potential only.
We calculate  $1/\tau_s$ and $1/\tau_g$ using Fermi golden rule, and
obtain the total scattering time as $1/\tau= 1/\tau_s + 1/\tau_g$:
\begin{align}
\frac{1}{\tau_s} &=\frac{2 \pi}{\hbar^2} \frac{N(E_F)}{4\pi^2}
\nonumber \\ &\times \int_0^\pi d\theta
\frac{1-\cos^2(\theta)}{2}\frac{\langle
V_s({\mathbf{q}})V_s({-\mathbf{q}})\rangle}{\epsilon^2(\mathbf{q})}|_{|\mathbf{q}|=2k_F\sin(\theta/2)}
\label{taus}
\end{align}
and
\begin{align}
\frac{1}{\tau_g} &=\frac{2 \pi}{\hbar^2} \frac{N(E_F)}{4\pi^2}
\nonumber \\ &\times \int_0^\pi d\theta [1-\cos(\theta)]\langle
\mathbf{A}_{\perp}({\mathbf{q}})\mathbf{A}_{\perp}({-\mathbf{q}})\rangle|_{|\mathbf{q}|=2k_F\sin(\theta/2)}
\label{taug},
\end{align}
where $\mathbf{A}_{\perp}(\mathbf{q})$ is the component of
$\mathbf{A}$ perpendicular to $\mathbf{q}$, $N(E_F)= \frac{k_F}{2\pi
\hbar v_F}$ is the one-valley density of states at the Fermi energy,
$\epsilon(\mathbf{q})= (\epsilon_0+1)/2+4e^2k_F/v_F|\mathbf{q}|$  is
the dielectric function including the substrate contribution, and
$k_F$, $v_F$, and $E_F$ are the Fermi momentum, velocity, and
energy. We extract the mobility from $\mu = \sigma/ne =
2\frac{e^2}{h}\frac{E_F \tau}{\hbar ne}$ (the factor of 2 accounts
for the two valleys). To calculate the magnitude of charge
fluctuations we use the relation $n(\mathbf{r}) = \frac{1}{\pi}
(\frac{V_s(\mathbf{r})}{\hbar v_F})^2$ between local charge density
and potential, from which:
\begin{equation}
n^*= \frac{1}{\pi} \frac{\langle V_s(\mathbf{r})^2\rangle }{(\hbar
v_F)^2}=\frac{1}{4\pi^3 \hbar^2 v_F^2} \int d^2\mathbf{q}\frac{\langle
V_s({\mathbf{q}})V_s({-\mathbf{q}})\rangle}{\epsilon^2(\mathbf{q})}.
\end{equation}

Since the correlation functions of all the potentials are determined
by the same correlation function describing the random strain field,
$\mu$ and $n^*$ are related. We find in all cases a linear relation
between $1/\mu$ and $n^*$ (within logarithmic corrections) with a
slope determined by the elastic coefficients of graphene, whose
specific expression differs for out-of-plane and for in-plane
strain. For out-of-plane strain we have
\begin{equation}
\frac{1}{\mu} =
n^*\frac{h}{4e}\left[\frac{\hbar^2v_F^2}{8e^4}+\frac{g_2^2(\lambda_L+\mu_L)^2}{g_1^2\mu_L^2}\right]
\frac{1}{\log[1/(k_F(n^*)a)]},
\end{equation}
whereas for in-plane strain we obtain
\begin{align}
 \frac{1}{\mu} &=  n^* \frac{h}{4e}
 \left[\frac{\hbar^2v_F^2}{16e^4} + \frac{ g_2^2}{g_1^2}
 \left( 1 + \frac{ ( \lambda_L + 2 \mu_L )^2}{\mu_L^2} \right) \right] \nonumber \\ &\times
 \frac{1}{\log[1/(k_F(n^*)a)]}.
 \label{munstar}
 \end{align}

In both expressions, the first term in the square bracket originates
from the contribution to scattering of the scalar potential and the
second from that of the pseudo-magnetic field. In these expressions,
$g_1$ and $g_2$ quantify the strength of electron-phonon coupling in
graphene, $\mu_L=9.4$ eV/\AA$^{2}$ and $\lambda_L=3.3$ eV/\AA$^{2}$
are Lam\'{e} coefficients~\cite{Zakharchenko2009}, and
$\frac{e^2}{\hbar v_F}=2.2$ ($a$ is the lattice constant of graphene
and the logarithm appears when cutting off the integrals at large
$q$-values, at $q=1/a$). Eqs. (5) and (6) show that the relation
between $1/\mu$ and $n^*$ is linear (the deviations caused by the
logarithm are within the fluctuations in the data, and in fact
improve the overall agreement) as found experimentally. Notably,
these relations only depend on fundamental constants and on the
elastic properties of graphene. In  this regard, the only role of
the substrate is to determine the magnitude of the strain present in
the graphene lattice.

The dashed lines in Fig. 3(a),(b) are best fits to the data
($\frac{1}{\mu} = \frac{h}{e}n^* \times 0.118$). Both expressions
above for random out-of-plane or in-plane strain reproduce this
value of the slope with realistic values of the $g_1$ and $g_2$
parameters (the slope only depends on their ratio). The parameter
$g_2$ is determined by the modulation of the hopping between $p_z$
orbitals and the strain, and it can be extracted from measurements
of effective magnetic fields created in highly strained
graphene~\cite{Letal1o}. A reasonable value is $g_2 \approx 2.5$
eV~\cite{GTGKP12}. The parameter $g_1$ gives the strength of the
scalar potential, and  estimates of its magnitude vary in the range
$g_1 \approx 4 - 10$ eV~\cite{OS66,SA02b,CJS10,GTGKP12}. Using
$n^*=10^{11} cm^{-2}$ and fixing $g_2 = 2.5$ eV, the expression for
random strain due to ripples Eq.(5) reproduces the slope of the
$1/\mu$-vs-$n^*$ for $g_1=3.65$ eV, and if Eq. (6) for in-plane
strain is taken, the experimental value is obtained for $g_1=6.9$
eV, in all cases fully compatible with the expected range of values.
We conclude that random strain quantitatively accounts for the
$1/\mu$-vs-$n^*$ relation observed in the experiments. While both
in-plane or out-of-plane random strain contribute, it is likely that
on hBN substrates in-plane strain dominates.

Having fixed the values of $g_2/g_1$ by comparing the theoretical
expression for $1/\mu$ with the experimental data, we can determine
whether it is the scalar or the gauge potential originating from
strain that gives the dominant contribution to the scattering time.
Interestingly we find that for both out-of-plane and in-plane random
strain, the scattering time associated to the random gauge potential
$\tau_g$ is approximately one order of magnitude smaller than the
scattering time associated to the scalar potential $\tau_s$, i.e. it
is the gauge potential that poses the most stringent limit to the
mobility. This is exactly what we would expect from our analysis of
weak-localization, and specifically from the experimental
observation that $\tau_*\simeq \tau$. This finding also explains why
the use of high dielectric constant substrates (such as SrTiO$_3$)
cannot lead to a major increase in mobility~\cite{Couto2011}: a
high-$\epsilon$ substrate could screen the deformation potential
--which is electrostatic in nature-- but not the effect of a random
pseudo-magnetic field. We conclude that our theoretical analysis of
the $1/\mu$-vs-$n^*$ relation does not only reproduce the
experimental data with realistic values of the model parameters, but
it is also internally consistent with other independent experimental
observations. It is this level of quantitative agreement and
internal consistency of results obtained by means of different
techniques that strongly supports the validity of our
interpretation.

\section{Conclusions}
The experimental and theoretical results discussed above lead to a
consistent physical scenario which can be understood only if random
strain fluctuations are the dominant source of disorder in graphene
on hBN (and other) substrates. We summarize the key points. The
analysis of weak-localization measurements shows that $\tau_{iv}\gg
\tau$, implying that scattering of charge carriers occurs mainly
within the same valley, and that is therefore due to a long-range
potential. It also shows that the characteristic time to break the
effective single-valley time reversal symmetry $\tau_*$ is
comparable to $\tau$, the elastic scattering time extracted from the
mobility, a finding that  can be explained naturally if random
pseudo-magnetic fields  due to strain are the dominant scattering
mechanism. Since this finding ($\tau_* \simeq \tau$) does not appear
to be compatible with any other disorder mechanism, the indication
that it provides as to the relevance of random strain fluctuations
is particularly compelling. The role of local strain fluctuations is
further confirmed by the correlation between the maximum observed
mobility with the line width of the Raman 2D-peak measured on the
very same devices (which has been identified as a measure of the
intensity of local mechanical deformations, i.e. local strain).
Finally, a conceptually straightforward theoretical analysis shows
that strain provides a qualitative and quantitative understanding of
the linear relation between $1/\mu$ and $n^*$. This same analysis
confirms that strain-induced disorder mainly generates scattering
through random pseudo-magnetic fields, and not through the scalar
deformation potential, which is precisely what we had concluded
independently through the study of weak localization.

Although most considerations above have been made for graphene on
hBN, our results point to the relevance of strain fluctuation also
for graphene on SiO$_2$ and SrTiO$_3$ substrates. Indeed, data
obtained from devices on SiO$_2$ and SrTiO$_3$ satisfy
quantitatively the same $1/\mu$-vs-$n^*$ relation that we have found
analyzing many devices on hBN. For graphene on SiO$_2$,
weak-localization measurements done in the
past~\cite{Tikhonenko2008,Guignard2012} allow us to draw conclusions
similar to those that we have discussed here for devices on hBN.
Additionally, random strain fluctuations explain why devices made on
substrates with extremely different surface chemistry show similar
mobility ($\approx 5.000-10.000$ cm$^2$/Vs), a fact that would be
difficult to understand if charge impurities at the substrate
surface were the dominating source of disorder (simply because the
density of charged impurities should depend very strongly on the
specific chemical groups present at the substrate surface). Finally,
the finding that strain fluctuations dominantly couples to the
electrons through the generation of a random pseudo-magnetic field
--and not through the deformation potential-- explains why the
mobility in devices on SrTiO$_3$ substrates~\cite{Couto2011}, which
have a very high dielectric constant, is not much higher than on
SiO$_2$, since the effect of magnetic field cannot be screened
electrostatically.

\section{Acknowledgements}
AFM gratefully acknowledges support by the SNF and by the NCCR QSIT.
FG acknowledges support from the Spanish Ministry of Economy
(MINECO) through Grant No. FIS2011-23713 and the European Research
Council Advanced Grant (contract 290846). CS and SE acknowledge
experimental help from F. Buckstegge, J. Dauber, B. Terr\'es, F.
Vollmer and M. Dr\"{o}geler and financial support from DFG and ERC
(contract 280140). AFM, FG and CS acknowledge funding from the EU
under the Graphene Flagship.

\section{Appendix A: Analysis of strain distributions.}

We discuss the technical details of the analysis of the effects of
random strain, and derive the expressions for the relations between
$1/\mu$ and $n^*$ reported in the main text. Strains can be induced
either by out-of-plane corrugations, or by in-plane displacements of
the atoms in the graphene lattice. We analyze the two cases
separately. We emphasize that this same analysis is consistent with
the observed density dependence of the mobility: irrespective of
whether strain is in-plane or out-of-plane, the calculated mobility
is independent of carrier density, within logarithmic corrections
that cause a slow mobility suppression at large $n$.

\subsection{Out of plane corrugations.}
We assume a given height profile, $h ( \vec{\bf r} )$. The height
corrugations lead to strains, which induce a scalar and a gauge
potential acting on the electrons~\cite{Vozmediano2010}:
\begin{align}
V_s ( \vec{\bf q} ) &= - g_1 \frac{\mu_L}{\lambda_L + 2 \mu_L}
\frac{q_x^2 + q_y^2}{\left| \vec{\bf q} \right|^4} {\cal F} \left(
\vec{\bf q} \right) \nonumber \\
A_x ( \vec{\bf q} ) &= g_2 \frac{\lambda_L + \mu_L}{\lambda_L + 2
\mu_L} \frac{q_x^2 - q_y^2}{\left| \vec{\bf q} \right|^4} {\cal F}
\left( \vec{\bf q} \right) \nonumber \\ A_y ( \vec{\bf q} ) &= - 2
g_2 \frac{\lambda_L + \mu_L}{\lambda_L + 2 \mu_L} \frac{q_x
q_y}{\left| \vec{\bf q} \right|^4} {\cal F} \left( \vec{\bf q}
\right) \label{potential}
\end{align}
where $g_1$ and $g_2$ are parameters with dimensions of energy,
$\lambda_L$ and $\mu_L$ are the elastic Lam\'e coefficients. ${\cal
F} ( \vec{\bf q} ) = \sum_{i,j} q_i q_j f_{i,j} ( \vec{\bf q} ) - |
\vec{\bf q} |^2 \sum_i f_{i,i} ( \vec{\bf q} )$, with $f_{i,j} (
\vec{\bf q} )$ the Fourier transform of $f_{i,j} ( \vec{ \bf r} ) =
\partial_i h ( \vec{\bf r} ) \partial_j  h ( \vec{ \bf r} )$,  $g_2
= 3 c \beta \gamma_0 / 4$, with $\gamma_0 \approx 2.7 {\rm eV},
\beta = \partial \log ( \gamma_0 ) / \partial \log ( a ) \approx 2$,
and $c = \mu_L / [ \sqrt{2} ( \lambda_L + \mu_L ) ] \approx
0.59$~\cite{Zakharchenko2009}.

We assume that the height correlations are such that
\begin{align}
\langle h ( \vec{\bf q} ) h ( - \vec{\bf q} ) \rangle &= \frac{A}{| \vec{\bf q} |^4}
\end{align}
where $A$ is a constant. This dependence corresponds to the profile
of a membrane with temperature $k_B T \propto A \kappa$, where
$\kappa$ is the bending rigidity of
graphene~\cite{katsnelson_corrug}. This assumption leads to $\langle
{\cal F} ( \vec{\bf q} ) {\cal F} ( - \vec{\bf q} ) \rangle =
\bar{A} | \vec{\bf q} |^2$, where $\bar{A}$ is a dimensionless
constant. It is given, approximately, by
\begin{align}
\bar{A} &\sim \frac{h_{r}^4}{\ell_{r}^4}
\end{align}
where $h_{r}$ and $\ell_{r}$ are typical values for the height and
size of the ripples. Using eqs. (\ref{taus}) and (\ref{taug}), we
find
\begin{align}
\tau_s^{-1} & \approx \left\{ \begin{array}{lr} \frac{ v_F g_1^2
\mu_L^2  \bar{A}}{32 \pi ( \lambda_L + 2 \mu_L )^2 e^4 k_F}  +
\cdots & \frac{\epsilon_0 \hbar v_F}{e^2} \lesssim 1 \\ \frac{g_1^2
\mu_L^2 \bar{A}}{8 \pi ( \lambda_L + 2 \mu_L )^2 (\epsilon_0+1)^2
\hbar^2 v_F k_F}  + \cdots &\frac{\epsilon_0 \hbar v_F}{e^2}
\gtrsim 1 \end{array} \right. \nonumber \\ \tau_g^{-1} &\approx
\frac{g_2^2 ( \lambda_L + \mu_L )^2 \bar{A}}{4 \pi \hbar^2 v_F k_F
(\lambda_L + 2 \mu_L )^2} \label{tau_1}
\end{align}
The mobility is given by $\mu = \sigma / ( n e ) = 2 \frac{e^2}{h}
\frac{v_F k_F \tau}{n e} =  \frac{2 e \pi}{h} \frac{v_F \tau}{k_F}$.
\\ For $\epsilon_0 \lesssim e^2 / ( \hbar v_F )$, the mobility is
\begin{align}
\frac{1}{\mu} &= \frac{\hbar \bar{A}}{e} \left[ \frac{g_1^2
\mu_L^2}{32 \pi e^4( \lambda_L + 2 \mu_L )^2} + \frac{g_2^2 (
\lambda_L + \mu_L )^2}{4 \pi ( \hbar v_F)^2 ( \lambda_L + 2 \mu_L
)^2} \right] \label{mu1}
\end{align}

The scalar potential in eq.(\ref{potential}) gives rise to charge
fluctuations, whose amplitude is given by:
\begin{widetext}
\begin{align}
n^* &= \frac{\langle V_s^2 ( \vec{\bf r} ) \rangle}{\pi \hbar^2
v_F^2} = \frac{1}{4 \pi^3 \hbar^2 v_F^2} \int d^2 \vec{\bf q}
\frac{\langle V_s ( \vec{\bf q} ) V_s ( - \vec{\bf q} )
\rangle}{\epsilon^2 ( \vec{\bf q} )} \approx \frac{g_1^2 \mu_L^2
\bar{A}}{2 \pi^2 ( \lambda_L + 2 \mu_L )^2 ( \hbar v_F)^2} \log
(\frac{1}{k_F(n^*)}). \label{n_star}
\end{align}
\end{widetext}
The ratio of the two expressions above leads to the $1/\mu$-vs-$n^*$
relation, Eq. (5) in the main text.

\subsection{In plane strains.}
A supporting substrate induces forces on the carbon atoms of a
graphene layer, leading to strains and deformations. Therefore, next
to out of plane deformations (discussed in the previous section)
that can occur on a corrugated substrate, or because of imperfect
adhesion during the graphene transfer process, in plane forces on
the Carbon atoms can also be expected. These forces induce strains,
which modify the electronic properties. In particular, periodic
interactions, associated to the incommensuration between the
lattices of graphene and the substrate, lead to the formation of
superstructures and Moir\'e patterns~\cite{Wetal14o,JDAM14,SGSG14}.
In addition, a random distribution of forces should be expected, due
to impurities in the substrates, and other imperfections in the
graphene/substrate system (e.g., remnants of adsorbates in between
the substarte and graphene).

We neglect the short range, periodic component of the interaction
potential between graphene and the substrate, and consider a random
potential, $V ( \vec{\bf r} )$, which varies slowly  over a distance
$\xi\gg a$, where $a$ is the lattice constant
\begin{align}
\left\langle V \left( \vec{\bf r} \right) V \left( \vec{\bf r}'
\right) \right\rangle &\approx \bar{V}^2 \xi^2 \delta \left(
\vec{\bf r} - \vec{\bf r}' \right)
\end{align}
This potential leads to forces at the positions of the carbon atoms
\begin{align}
\vec{\bf F} \left( \vec{\bf r} \right) &= \nabla V \left( \vec{\bf
r} \right) \label{forces}
\end{align}
The elastic energy of the graphene lattice is
\begin{align}
{\cal H}_{elastic} &= \frac{\lambda}{2} \int d^2 \vec{\bf r}  \left(
\sum_{i=x,y} u_{ii} \right)^2 + \mu \int d^2 \vec{\bf r}
\sum_{i,j=x,y} u_{ij}^2 \nonumber \\ &+ \int \frac{d^2 \vec{\bf
r}}{A} \vec{\bf F} \left( \vec{\bf r} \right) \vec{u} \left(
\vec{\bf r} \right) \label{e_elastic}
\end{align}
where $A = \sqrt{3} d_G^2 / 2$ is the area of the unit cell, $d_G$
is the lattice constant, $u \left( \vec{\bf r} \right)$ is the
displacement of the atom at position $\vec{\bf r}$ from its
equilibrium position, $u_{ij} = \left( \partial_i u_j + \partial_j
u_i \right) / 2 $, and we assume that the displacements are small,
so that the assumption of a linear coupling to local forces is
valid.

We Fourier transform eq. (\ref{e_elastic})
\begin{align}
{\cal H}_{elastic} &= \frac{\lambda}{2} \sum_{\vec{\bf k}} \left(
\vec{\bf k} \vec{\bf u}_{\vec{\bf k}} \right)^2  + \mu
\sum_{\vec{\bf k}} \frac{\left( k_i u_j + k_j u_i \right)^2}{4}
\nonumber \\ &+ \sum_{\vec{\bf k}} \frac{\vec{\bf F}_{\vec{\bf k}}
\vec{\bf u}_{\vec{\bf k}}}{A}
\end{align}
For long wavelength force distributions, $\left| \vec{\bf k} \right|
\ll \left| \vec{\bf G} \right|$, where $\vec{\bf G}$ is a reciprocal
vector of the graphene lattice, the displacements are
\begin{align}
\vec{\bf u}_{\vec{\bf k}} &= - \frac{\vec{\bf F}_{\vec{\bf
k}}^\parallel}{A ( \lambda + 2 \mu ) \left| \vec{\bf k} \right|^2} -
\frac{\vec{\bf F}_{\vec{\bf k}}^\perp}{A \mu \left| \vec{\bf k}
\right|^2}
\end{align}
where the $\parallel$ and $\perp$ superscripts stand for the
parallel and transverse components of $\vec{\bf F}_{\vec{\bf
 k}}$ with respect to $\vec{\bf k}$ ($\vec{\bf F}_{\vec{\bf k}} = i \vec{\bf k}
V_{\vec{\bf k}}$, the vector $\vec{\bf F}_{\vec{\bf k}}$ has only a
longitudinal component, but this is not the generic case, see
below.). If $\left| \vec{\bf k} - \vec{\bf G} \right| = \left|
\delta \vec{\bf k}\right| \ll \left| \vec{\bf G} \right|$, long
wavelength displacements are also generated
\begin{align}
\vec{\bf u}_{\delta \vec{\bf k}} &= - \frac{\vec{\bf F}_{\vec{\bf
k}}^\parallel}{ A( \lambda + 2 \mu ) \left| \delta \vec{\bf k}
\right|^2} - \frac{\vec{\bf F}_{\vec{\bf k}}^\perp}{A \mu \left|
\delta \vec{\bf k} \right|^2}
\end{align}
where now the superscripts $\parallel$ and $\perp$ refer to the
orientation of $\vec{\bf F}_{\vec{\bf k}}$ with respect to $\delta
\vec{\bf k}$. From $\vec{\bf u}_{\vec{\bf k}}$ we can obtain the
strain tensor
\begin{align}
u_{i,j} \left( \vec{\bf k} \right) &= \frac{k_i \left. \vec{\bf
u}_{\vec{\bf k}} \right|_j + k_j \left. \vec{\bf u}_{\vec{\bf k}}
\right|_i}{2}
\end{align}
For $| \vec{\bf k} | \ll | \vec{\bf G} |$ we have $\vec{\bf
F}_{\vec{\bf k}} = \vec{\bf k} V_{\vec{\bf k}}$, while for $| \delta
\vec{\bf k} | = | \vec{\bf k} - \vec{\bf G} | \ll | \vec{\bf G} |$
we obtain $\vec{\bf F}_{\delta \vec{\bf k}} \approx \vec{\bf G}
V_{\vec{\bf G}}$. Thus, we obtain two different contributions to the
correlations of long range strains
\begin{align}
\left\langle u_{i,j} ( \vec{\bf k} ) u_{i,j} ( - \vec{\bf k} )
\right\rangle &\propto \left\langle V ( \vec{\bf k} ) V ( - \vec{\bf
k} ) \right\rangle
\nonumber \\
\left\langle u_{i,j} ( \delta \vec{\bf k} ) u_{i,j} ( - \delta
\vec{\bf k} ) \right\rangle &\propto \frac{| \vec{\bf G} |^2}{|
\delta \vec{\bf k} |^2} \left\langle V ( \vec{\bf G} ) V ( -
\vec{\bf G} ) \right\rangle \label{strains}
\end{align}
where $\left\langle V ( \vec{\bf k} ) V ( - \vec{\bf k} )
\right\rangle = \bar{V}^2 \xi^2$. In the following we will
concentrate on the effects of the random components of the potential
of order $| \vec{\bf G} | = ( 4 \pi ) / (\sqrt{3} d_G)$. The
contribution to the transport properties is larger that that from
small momenta by terms of order $| \vec{\bf G} |^2/k_F^2$.

The scalar and gauge potentials are
\begin{align}
V_s ( \delta \vec{\bf k} ) &=  g_1 \sum_{\vec{\bf G}} \frac{( G_x
\delta k_x + G_y \delta k_y ) V_{\vec{\bf k}}}{( \lambda_L +2 \mu_L
) |
\delta \vec{\bf k} |^2} \nonumber \\
A_x ( \delta \vec{\bf k} ) & = \frac{g_2}{\hbar v_F} \sum_{\vec{\bf
G}}
\frac{ V_{\vec{\bf k}}}{| \delta \vec{\bf k} |^4} \nonumber \\
&\times \left[ \left[ ( \delta k_x )^2 - ( \delta k_y )^2 \right]
\frac{G_x \delta k_x + G_y \delta k_y} {\lambda_L + 2 \mu_L} \right.
\nonumber \\ &\left. + ( 2 \delta k_x \delta k_y ) \frac{- G_x
\delta k_y + G_y \delta k_x}
{\mu_L} \right] \nonumber \\ \nonumber \\
A_y ( \delta \vec{\bf k} ) & = \frac{g_2´}{\hbar v_F}
\sum_{\vec{\bf G}}
\frac{  V_{\vec{\bf k}}}{| \delta \vec{\bf k} |^4} \nonumber \\
&\times \left[ - ( 2 \delta k_x \delta k_y ) \frac{ G_x \delta k_x +
G_y \delta k_y} {\lambda_L + 2 \mu_L} \right. \nonumber \\ &+ \left.
\left[ ( \delta k_x )^2 - ( \delta k_y )^2 \right] \frac{- G_x
\delta k_y + G_y \delta k_x} {\mu_L} \right] \label{gauge}
\end{align}
The gauge potential can be divided into a component $\vec{\bf
A}_\parallel$, parallel to $\delta \vec{\bf k}$, and a component
$\vec{\bf A}_\perp$, perpendicular to $\delta \vec{\bf k}$. The
effect of $\vec{\bf A}_\parallel$ can be gauged away, and only
$\vec{\bf A}_\perp$ gives a physical effect. In terms of the
potential correlations we find
\begin{align}
\langle V_s ( \delta \vec{\bf k} ) V_s ( - \delta \vec{\bf k} )
\rangle &= \sum_{\vec{\bf G}} \frac{g_1^2 | \vec{\bf G} |^2
\bar{V}^2 \xi^2 \cos^2 ( \theta )} {( \lambda_L + 2 \mu_L )^2
| \delta \vec{\bf k} |^2 A^2}  \nonumber \\
\langle \vec{\bf A}_\perp ( \delta \vec{\bf k} ) \vec{\bf A}_\perp (
- \delta \vec{\bf k} ) \rangle &= \sum_{\vec{\bf G}} \frac{g_2^2 |
\vec{\bf G} |^2 \bar{V}^2 \xi^2  }{( \hbar v_F )^2 | \delta \vec{\bf
k} |^2 A^2} \nonumber \\ &\times \left[ \frac{\sin^2 ( 3 \theta )
\cos^2 ( \theta )}{( \lambda_L + 2 \mu_L )^2} + \frac{\cos^2 ( 3
\theta ) \sin^2 ( \theta )}{\mu_L^2} \right] \label{corr}
\end{align}
where $\theta$ is the angle between $\vec{\bf G}$ and $\delta
\vec{\bf k}$, $A = d_G^2 \sqrt{3} / 2$ is the area of the unit cell
of the graphene lattice, and we neglect terms proportional to
quantities like $\sin ( \theta ) \cos ( \theta)$ which average to
zero when summing over $\vec{\bf G}$. The factor $\cos ( 3 \theta )$
arises from extracting the component of $\vec{\bf A}$ normal to
$\delta \vec{\bf k}$.

Using eqs. (\ref{taus}) and (\ref{taug}), we obtain
\begin{align}
\frac{\hbar}{\tau_s} &\approx \frac{\pi g_1^2 \bar{V}^2 \xi^2}{6 ( \lambda_L + 2 \mu_L )^2 d_G^6 \alpha^2 \hbar v_F k_F} \nonumber \\
\frac{\hbar}{\tau_g} & \approx \frac{8 \pi g_2^2 \bar{V}^2 \xi^2}{3
d_G^6 \hbar v_F k_F} \left[ \frac{1}{( \lambda_L + 2 \mu_L )^2} +
\frac{1}{\mu_L^2} \right] \label{tau}
\end{align}
The mobility is
\begin{align}
\frac{1}{\mu} &= \frac{e n}{\sigma} = \frac{h}{e} \frac{k_F}{2 \pi
v_F} \left( \frac{1}{\tau_s} + \frac{1}{\tau_g} \right)
\label{mobility}
\end{align}
The carrier fluctuations at the neutrality point are given by
\begin{align}
n^* &\approx \frac{16 g_1^2 \bar{V}^2 \xi^2}{3 ( \hbar v_F )^2 (
\lambda_L + 2 \mu_L )^2 d_G^6} \log \left( \frac{1}{k_F(n^*)a}
\right) \label{nstar}
\end{align}
From the ration of the last two expression we obtain Eq.(6) in the
main text.

\section*{APPENDIX B: 1/$\mu$-vs-$n^*$ RELATION FOR BILAYER GRAPHENE}
\begin{figure}[t]
\includegraphics[width=0.4\textwidth]{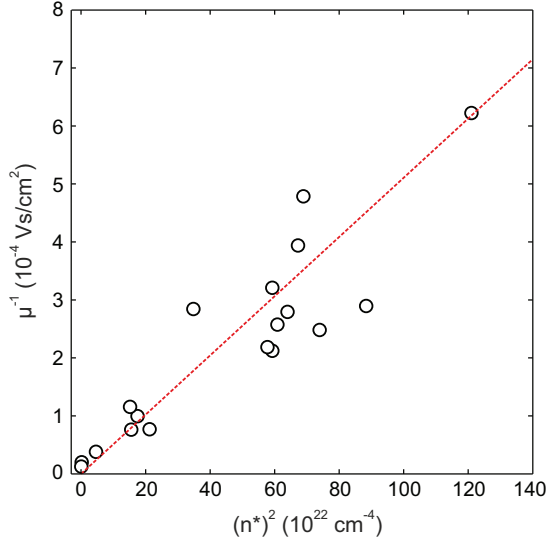}
\caption{(color online) Correlation of the inverse mobility and
$(n^*)^2$ for a number of different bilayer graphene flakes on
different substrates. } \label{bilayerfig}
\end{figure}

Here, we illustrate that the linearity of the relation between
$\mu^{-1}$ and $n^*$ for monolayer graphene is not trivial, by
comparing the result shown in Fig. 3 to those of a similar analysis
done for bilayer graphene devices (the bilayer devices have been
fabricated on hBN and SiO$_2$ substrates, following protocols
identical to those used for the monolayers). The result is
illustrated in Fig.~\ref{bilayerfig}, in which we plot $\mu^{-1}$ as
a function of $(n^*)^2$: it is apparent that the experimental data
obey a linear relation, i.e., for bilayers $\mu^{-1} \propto
(n^*)^2$, in contrast to the relation $\mu^{-1} \propto (n^*)$ found
in monolayers. This finding further supports our theoretical
analysis of disorder in monolayers in terms of strain, which
correctly captures the observed, non-trivial linear dependence of
the relation between $1/\mu$ and $n^*$.

\section{Appendix C: Weak-Localization and single-valley effective time reversal symmetry}
Weak-localization measurements in graphene provide a wealth of
information about the scattering processes that take place in the
material. As we have discussed in the main text, we can conclude
directly from the results of the fits of the magneto-resistance
curves that the intervalley scattering time $\tau_{iv}$ is much
longer than the elastic scattering time $\tau$, which indicates that
intra-valley scattering processes --and therefore long-range
potentials-- dominate the effect of disorder. We have also found
that the elastic scattering time $\tau$ nearly coincides with the
time $\tau_{*}$ needed to break the effective single-valley time
reversal symmetry, and argued that this observation strongly
indicates that strain --and not charged impurities-- is the dominant
scattering source. As the reader may not be fully familiar with the
concept of effective single-valley time reversal symmetry, we
discuss this here in some more detail for completeness (for more
information, see Ref.~\cite{Morpurgo2006}).

Effective single-valley time reversal symmetry is a concept relevant
for graphene, in the regime in which a continuum Dirac Hamiltonian
provides a good description (i.e., when the Fermi level is not too
far away from the charge neutrality point). In the ideal case, the
Dirac Hamiltonian (where $\mathbf{k}$ is the momentum relative to
the K-point)
\begin{equation}
H= \hbar v_F \left(
  \begin{array}{cc}
    0 & k_x + i k_y \\
    k_x - i k_y & 0 \\
  \end{array}
\right)
\end{equation}
is invariant upon the anti-unitary transformation
$i\hat{\sigma_y}\hat{K}$, with $\hat{K}$ denoting complex
conjugation. This anti-unitary transformation mimics the
implementation of time-reversal symmetry, as --for each electronic
state-- it sends $\mathbf{k}$ into $-\mathbf{k}$ and reverses the
spin. However, this is not the true time reversal symmetry
operation. Indeed, true time reversal symmetry changes the sign of
the total momentum (and not just of the momentum relative to the K
point) and sends states in one of the valley into the other valley
(time reversal symmetry therefore cannot be implemented by
considering one valley only). That is why $i\hat{\sigma_y}\hat{K}$
is referred to as "effective single-valley time reversal symmetry".
While it remains a good symmetry as long as $H$ is well approximated
by the Dirac Hamiltonian, the implication of not being the "true"
time reversal operation is that it can be easily violated, by
different microscopic mechanisms.

For instance, effective time reversal symmetry is violated by the
quadratic momentum terms that are neglected when making the linear
approximation in the continuum, which leads to the Dirac
Hamiltonian. It is obvious that --being quadratic-- these terms do
not change sign upon inverting the sign of $\mathbf{k}$, whereas the
linear terms in the Dirac Hamiltonian do. As a result, when
including both the linear and quadratic terms in $\mathbf{k}$, the
single-valley Hamiltonian is not invariant upon effective
single-valley time reversal symmetry. This mechanism, however,
cannot account for our experimental observations ($\tau_* \simeq
\tau$ independent of carrier density): the effect of the quadratic
terms become more relevant as $E_F$ is increased further away from
the charge neutrality point. If these terms were the relevant ones
in determining the characteristic time scale $\tau_{*}$, we should
find that $\tau_{*}$ becomes shorter at larger carrier density,
contrary to what we observe experimentally (see Fig. 4c of the main
text). Additionally, this mechanism cannot explain why $\tau_*
\simeq \tau$, because the quadratic terms in $\mathbf{k}$ do not
cause any scattering of electron waves.

Other two mechanisms that break effective time reversal symmetry are
strain, and the presence of a "gap" term in the Dirac Hamiltonian.
Strain breaks the effective time reversal symmetry by generating a
random  pseudo-magnetic field. Indeed, within a single valley, this
pseudo magnetic field acts on the orbital degrees of freedom in all
regards as a true magnetic field, i.e., it is described by a gauge
potential minimally coupled to the momentum. If the dominant source
of scattering are spatial inhomogeneities in this gauge potential,
such a mechanism very naturally explains why the elastic scattering
time $\tau$ and the time needed to break the effective time reversal
symmetry $\tau_{*}$ nearly coincide, as scattering and effective
time reversal symmetry breaking originate from the same term in the
Hamiltonian.

A "gap" term --i.e., a difference $\Delta$ in on-site energy between
the A and B carbon atom in the unit cell of graphene-- also breaks
the effective single-valley time reversal symmetry. In that case the
Hamiltonian becomes
\begin{equation}
H=  \left(
  \begin{array}{cc}
    \Delta/2 & \hbar v_F(k_x + i k_y) \\
    \hbar v_F(k_x - i k_y) & -\Delta/2 \\
  \end{array}
\right)
\end{equation}
The fact that this Hamiltonian is not invariant upon
$i\hat{\sigma_y}\hat{K}$ can be checked by a direct calculation. It
is also obvious without doing any calculation, if we observe that
this Hamiltonian is formally identical to that of Rashba spin-orbit
interaction in the presence of a Zeeman term (with $g\mu B$
corresponding to $\Delta$/2), a system which lacks time reversal
symmetry.

In our experimental case, charged impurities on a substrate can
contribute to the breaking of effective time reversal symmetry
through this mechanism. More specifically, charge impurities would
generate random electrostatic potentials. On average, these
potentials would be the same on the A and B atoms forming graphene.
Nevertheless, fluctuations would  exist so that locally the
electrostatic on-site energy would be slightly different on the A
and B atom in each unit cell, i.e., locally a non-zero $\Delta$ term
would be present. However, in our experiments, this mechanism cannot
explain why the scattering time $\tau$ and the characteristic time
for breaking effective time reversal symmetry $\tau_{*}$ would
coincide. In fact, as discussed in the main text, this mechanism
would predict that $\tau_{*}\gg \tau$, because $\tau$ is determined
by the Fourier components of the potential at $2k_F$, whereas
$\Delta$ is determined by the components at $k \simeq 1/a$ (with a
long ranged potential the latter are much smaller).


\begin{thebibliography}{52}%
\makeatletter
\providecommand \@ifxundefined [1]{%
 \@ifx{#1\undefined}
}%
\providecommand \@ifnum [1]{%
 \ifnum #1\expandafter \@firstoftwo
 \else \expandafter \@secondoftwo
 \fi
}%
\providecommand \@ifx [1]{%
 \ifx #1\expandafter \@firstoftwo
 \else \expandafter \@secondoftwo
 \fi
}%
\providecommand \natexlab [1]{#1}%
\providecommand \enquote  [1]{\textit{#1}}%
\providecommand \bibnamefont  [1]{#1}%
\providecommand \bibfnamefont [1]{#1}%
\providecommand \citenamefont [1]{#1}%
\providecommand \href@noop [0]{\@secondoftwo}%
\providecommand \href [0]{\begingroup \@sanitize@url \@href}%
\providecommand \@href[1]{\@@startlink{#1}\@@href}%
\providecommand \@@href[1]{\endgroup#1\@@endlink}%
\providecommand \@sanitize@url [0]{\catcode `\\12\catcode
`\$12\catcode
  `\&12\catcode `\#12\catcode `\^12\catcode `\_12\catcode `\%12\relax}%
\providecommand \@@startlink[1]{}%
\providecommand \@@endlink[0]{}%
\providecommand \url  [0]{\begingroup\@sanitize@url \@url }%
\providecommand \@url [1]{\endgroup\@href {#1}{\urlprefix }}%
\providecommand \urlprefix  [0]{URL }%
\providecommand \Eprint [0]{\href }%
\providecommand \doibase [0]{http://dx.doi.org/}%
\providecommand \selectlanguage [0]{\@gobble}%
\providecommand \bibinfo  [0]{\@secondoftwo}%
\providecommand \bibfield  [0]{\@secondoftwo}%
\providecommand \translation [1]{[#1]}%
\providecommand \BibitemOpen [0]{}%
\providecommand \bibitemStop [0]{}%
\providecommand \bibitemNoStop [0]{.\EOS\space}%
\providecommand \EOS [0]{\spacefactor3000\relax}%
\providecommand \BibitemShut  [1]{\csname bibitem#1\endcsname}%
\let\auto@bib@innerbib\@empty
\bibitem [{\citenamefont {Dean}\ \emph {et~al.}(2010)\citenamefont {Dean},
  \citenamefont {Young}, \citenamefont {Meric}, \citenamefont {Lee},
  \citenamefont {Wang}, \citenamefont {Sorgenfrei}, \citenamefont {Watanabe},
  \citenamefont {Taniguchi}, \citenamefont {Kim}, \citenamefont {Shepard},\
  and\ \citenamefont {Hone}}]{Dean2010}%
  \BibitemOpen
  \bibfield  {author} {\bibinfo {author} {\bibfnamefont {C.~R.}\ \bibnamefont
  {Dean}}, \bibinfo {author} {\bibfnamefont {A.~F.}\ \bibnamefont {Young}},
  \bibinfo {author} {\bibfnamefont {I.}~\bibnamefont {Meric}}, \bibinfo
  {author} {\bibfnamefont {C.}~\bibnamefont {Lee}}, \bibinfo {author}
  {\bibfnamefont {L.}~\bibnamefont {Wang}}, \bibinfo {author} {\bibfnamefont
  {S.}~\bibnamefont {Sorgenfrei}}, \bibinfo {author} {\bibfnamefont
  {K.}~\bibnamefont {Watanabe}}, \bibinfo {author} {\bibfnamefont
  {T.}~\bibnamefont {Taniguchi}}, \bibinfo {author} {\bibfnamefont
  {P.}~\bibnamefont {Kim}}, \bibinfo {author} {\bibfnamefont {K.~L.}\
  \bibnamefont {Shepard}}, \ and\ \bibinfo {author} {\bibfnamefont
  {J.}~\bibnamefont {Hone}},\ }\bibfield  {title} {\enquote {\bibinfo {title}
  {Boron nitride substrates for high-quality graphene electronics},}\
  }\href@noop {} {\bibfield  {journal} {\bibinfo  {journal} {Nat. Nano.}\
  }\textbf {\bibinfo {volume} {5}},\ \bibinfo {pages} {722--726} (\bibinfo
  {year} {2010})}\BibitemShut {NoStop}%
\bibitem [{\citenamefont {Zomer}\ \emph {et~al.}(2011)\citenamefont {Zomer},
  \citenamefont {Dash}, \citenamefont {Tombros},\ and\ \citenamefont {van
  Wees}}]{Zomer2011_TranshBN}%
  \BibitemOpen
  \bibfield  {author} {\bibinfo {author} {\bibfnamefont {P.~J.}\ \bibnamefont
  {Zomer}}, \bibinfo {author} {\bibfnamefont {S.~P.}\ \bibnamefont {Dash}},
  \bibinfo {author} {\bibfnamefont {N.}~\bibnamefont {Tombros}}, \ and\
  \bibinfo {author} {\bibfnamefont {B.~J.}\ \bibnamefont {van Wees}},\
  }\bibfield  {title} {\enquote {\bibinfo {title} {A transfer technique for
  high mobility graphene devices on commercially available hexagonal boron
  nitride},}\ }\href {\doibase dx.doi.org/10.1063/1.3665405} {\bibfield
  {journal} {\bibinfo  {journal} {Appl. Phys. Lett.}\ }\textbf {\bibinfo
  {volume} {99}},\ \bibinfo {eid} {232104} (\bibinfo {year}
  {2011})}\BibitemShut {NoStop}%
\bibitem [{\citenamefont {Mayorov}\ \emph {et~al.}(2011)\citenamefont
  {Mayorov}, \citenamefont {Gorbachev}, \citenamefont {Morozov}, \citenamefont
  {Britnell}, \citenamefont {Jalil}, \citenamefont {Ponomarenko}, \citenamefont
  {Blake}, \citenamefont {Novoselov}, \citenamefont {Watanabe}, \citenamefont
  {Taniguchi},\ and\ \citenamefont {Geim}}]{Mayorov2011_micron}%
  \BibitemOpen
  \bibfield  {author} {\bibinfo {author} {\bibfnamefont {A.~S.}\ \bibnamefont
  {Mayorov}}, \bibinfo {author} {\bibfnamefont {R.~V.}\ \bibnamefont
  {Gorbachev}}, \bibinfo {author} {\bibfnamefont {S.~V.}\ \bibnamefont
  {Morozov}}, \bibinfo {author} {\bibfnamefont {L.}~\bibnamefont {Britnell}},
  \bibinfo {author} {\bibfnamefont {R.}~\bibnamefont {Jalil}}, \bibinfo
  {author} {\bibfnamefont {L.~A.}\ \bibnamefont {Ponomarenko}}, \bibinfo
  {author} {\bibfnamefont {P.}~\bibnamefont {Blake}}, \bibinfo {author}
  {\bibfnamefont {K.~S.}\ \bibnamefont {Novoselov}}, \bibinfo {author}
  {\bibfnamefont {K.}~\bibnamefont {Watanabe}}, \bibinfo {author}
  {\bibfnamefont {T.}~\bibnamefont {Taniguchi}}, \ and\ \bibinfo {author}
  {\bibfnamefont {A.~K.}\ \bibnamefont {Geim}},\ }\bibfield  {title} {\enquote
  {\bibinfo {title} {Micrometer-scale ballistic transport in encapsulated
  graphene at room temperature},}\ }\href {\doibase 10.1021/nl200758b}
  {\bibfield  {journal} {\bibinfo  {journal} {Nano Lett.}\ }\textbf {\bibinfo
  {volume} {11}},\ \bibinfo {pages} {2396--2399} (\bibinfo {year}
  {2011})}\BibitemShut {NoStop}%
\bibitem [{\citenamefont {Wang}\ \emph {et~al.}(2013)\citenamefont {Wang},
  \citenamefont {Meric}, \citenamefont {Huang}, \citenamefont {Gao},
  \citenamefont {Gao}, \citenamefont {Tran}, \citenamefont {Taniguchi},
  \citenamefont {Watanabe}, \citenamefont {Campos}, \citenamefont {Muller},
  \citenamefont {Guo}, \citenamefont {Kim}, \citenamefont {Hone}, \citenamefont
  {Shepard},\ and\ \citenamefont {Dean}}]{Wang2013}%
  \BibitemOpen
  \bibfield  {author} {\bibinfo {author} {\bibfnamefont {L.}~\bibnamefont
  {Wang}}, \bibinfo {author} {\bibfnamefont {I.}~\bibnamefont {Meric}},
  \bibinfo {author} {\bibfnamefont {P.~Y.}\ \bibnamefont {Huang}}, \bibinfo
  {author} {\bibfnamefont {Q.}~\bibnamefont {Gao}}, \bibinfo {author}
  {\bibfnamefont {Y.}~\bibnamefont {Gao}}, \bibinfo {author} {\bibfnamefont
  {H.}~\bibnamefont {Tran}}, \bibinfo {author} {\bibfnamefont {T.}~\bibnamefont
  {Taniguchi}}, \bibinfo {author} {\bibfnamefont {K.}~\bibnamefont {Watanabe}},
  \bibinfo {author} {\bibfnamefont {L.~M.}\ \bibnamefont {Campos}}, \bibinfo
  {author} {\bibfnamefont {D.~A.}\ \bibnamefont {Muller}}, \bibinfo {author}
  {\bibfnamefont {J.}~\bibnamefont {Guo}}, \bibinfo {author} {\bibfnamefont
  {P.}~\bibnamefont {Kim}}, \bibinfo {author} {\bibfnamefont {J.}~\bibnamefont
  {Hone}}, \bibinfo {author} {\bibfnamefont {K.~L.}\ \bibnamefont {Shepard}}, \
  and\ \bibinfo {author} {\bibfnamefont {C.~R.}\ \bibnamefont {Dean}},\
  }\bibfield  {title} {\enquote {\bibinfo {title} {One-dimensional electrical
  contact to a two-dimensional material},}\ }\href {\doibase
  10.1126/science.1244358} {\bibfield  {journal} {\bibinfo  {journal}
  {Science}\ }\textbf {\bibinfo {volume} {342}},\ \bibinfo {pages} {614--617}
  (\bibinfo {year} {2013})}\BibitemShut {NoStop}%
\bibitem [{\citenamefont {Dean}\ \emph {et~al.}(2011)\citenamefont {Dean},
  \citenamefont {Young}, \citenamefont {Cadden-Zimansky}, \citenamefont {Wang},
  \citenamefont {Ren}, \citenamefont {Watanabe}, \citenamefont {Taniguchi},
  \citenamefont {Kim}, \citenamefont {Hone},\ and\ \citenamefont
  {Shepard}}]{dean2011_FQHE}%
  \BibitemOpen
  \bibfield  {author} {\bibinfo {author} {\bibfnamefont {C.~R.}\ \bibnamefont
  {Dean}}, \bibinfo {author} {\bibfnamefont {A.~F.}\ \bibnamefont {Young}},
  \bibinfo {author} {\bibfnamefont {P.}~\bibnamefont {Cadden-Zimansky}},
  \bibinfo {author} {\bibfnamefont {L.}~\bibnamefont {Wang}}, \bibinfo {author}
  {\bibfnamefont {H.}~\bibnamefont {Ren}}, \bibinfo {author} {\bibfnamefont
  {K.}~\bibnamefont {Watanabe}}, \bibinfo {author} {\bibfnamefont
  {T.}~\bibnamefont {Taniguchi}}, \bibinfo {author} {\bibfnamefont
  {P.}~\bibnamefont {Kim}}, \bibinfo {author} {\bibfnamefont {J.}~\bibnamefont
  {Hone}}, \ and\ \bibinfo {author} {\bibfnamefont {K.~L.}\ \bibnamefont
  {Shepard}},\ }\bibfield  {title} {\enquote {\bibinfo {title} {Multicomponent
  fractional quantum hall effect in graphene},}\ }\href@noop {} {\bibfield
  {journal} {\bibinfo  {journal} {Nat. Phys.}\ }\textbf {\bibinfo {volume}
  {7}},\ \bibinfo {pages} {693--696} (\bibinfo {year} {2011})}\BibitemShut
  {NoStop}%
\bibitem [{\citenamefont {Young}\ \emph {et~al.}(2012)\citenamefont {Young},
  \citenamefont {Dean}, \citenamefont {Wang}, \citenamefont {Ren},
  \citenamefont {Cadden-Zimansky}, \citenamefont {Watanabe}, \citenamefont
  {Taniguchi}, \citenamefont {Hone}, \citenamefont {Shepard},\ and\
  \citenamefont {Kim}}]{Young2012_svqHF}%
  \BibitemOpen
  \bibfield  {author} {\bibinfo {author} {\bibfnamefont {A.~F.}\ \bibnamefont
  {Young}}, \bibinfo {author} {\bibfnamefont {C.~R.}\ \bibnamefont {Dean}},
  \bibinfo {author} {\bibfnamefont {L.}~\bibnamefont {Wang}}, \bibinfo {author}
  {\bibfnamefont {H.}~\bibnamefont {Ren}}, \bibinfo {author} {\bibfnamefont
  {P.}~\bibnamefont {Cadden-Zimansky}}, \bibinfo {author} {\bibfnamefont
  {K.}~\bibnamefont {Watanabe}}, \bibinfo {author} {\bibfnamefont
  {T.}~\bibnamefont {Taniguchi}}, \bibinfo {author} {\bibfnamefont
  {J.}~\bibnamefont {Hone}}, \bibinfo {author} {\bibfnamefont {K.~L.}\
  \bibnamefont {Shepard}}, \ and\ \bibinfo {author} {\bibfnamefont
  {P.}~\bibnamefont {Kim}},\ }\bibfield  {title} {\enquote {\bibinfo {title}
  {Spin and valley quantum hall ferromagnetism in graphene},}\ }\href@noop {}
  {\bibfield  {journal} {\bibinfo  {journal} {Nat. Phys.}\ }\textbf {\bibinfo
  {volume} {8}},\ \bibinfo {pages} {550--556} (\bibinfo {year}
  {2012})}\BibitemShut {NoStop}%
\bibitem [{\citenamefont {Ponomarenko}\ \emph {et~al.}(2013)\citenamefont
  {Ponomarenko}, \citenamefont {Gorbachev}, \citenamefont {Yu}, \citenamefont
  {Elias}, \citenamefont {Jalil}, \citenamefont {Patel}, \citenamefont
  {Mishchenko}, \citenamefont {Mayorov}, \citenamefont {Woods}, \citenamefont
  {Wallbank}, \citenamefont {Mucha-Kruczynski}, \citenamefont {Piot},
  \citenamefont {Potemski}, \citenamefont {Grigorieva}, \citenamefont
  {Novoselov}, \citenamefont {Guinea}, \citenamefont {Fal'ko},\ and\
  \citenamefont {Geim}}]{Ponomarenko2013_Cloning}%
  \BibitemOpen
  \bibfield  {author} {\bibinfo {author} {\bibfnamefont {L.~A.}\ \bibnamefont
  {Ponomarenko}}, \bibinfo {author} {\bibfnamefont {R.~V.}\ \bibnamefont
  {Gorbachev}}, \bibinfo {author} {\bibfnamefont {G.~L.}\ \bibnamefont {Yu}},
  \bibinfo {author} {\bibfnamefont {D.~C.}\ \bibnamefont {Elias}}, \bibinfo
  {author} {\bibfnamefont {R.}~\bibnamefont {Jalil}}, \bibinfo {author}
  {\bibfnamefont {A.~A.}\ \bibnamefont {Patel}}, \bibinfo {author}
  {\bibfnamefont {A.}~\bibnamefont {Mishchenko}}, \bibinfo {author}
  {\bibfnamefont {A.~S.}\ \bibnamefont {Mayorov}}, \bibinfo {author}
  {\bibfnamefont {C.~R.}\ \bibnamefont {Woods}}, \bibinfo {author}
  {\bibfnamefont {J.~R.}\ \bibnamefont {Wallbank}}, \bibinfo {author}
  {\bibfnamefont {M.}~\bibnamefont {Mucha-Kruczynski}}, \bibinfo {author}
  {\bibfnamefont {B.~A.}\ \bibnamefont {Piot}}, \bibinfo {author}
  {\bibfnamefont {M.}~\bibnamefont {Potemski}}, \bibinfo {author}
  {\bibfnamefont {I.~V.}\ \bibnamefont {Grigorieva}}, \bibinfo {author}
  {\bibfnamefont {K.~S.}\ \bibnamefont {Novoselov}}, \bibinfo {author}
  {\bibfnamefont {F.}~\bibnamefont {Guinea}}, \bibinfo {author} {\bibfnamefont
  {V.~I.}\ \bibnamefont {Fal'ko}}, \ and\ \bibinfo {author} {\bibfnamefont
  {A.~K.}\ \bibnamefont {Geim}},\ }\bibfield  {title} {\enquote {\bibinfo
  {title} {Cloning of dirac fermions in graphene superlattices},}\ }\href
  {\doibase 10.1038/nature12187} {\bibfield  {journal} {\bibinfo  {journal}
  {Nature}\ }\textbf {\bibinfo {volume} {497}},\ \bibinfo {pages} {594--597}
  (\bibinfo {year} {2013})}\BibitemShut {NoStop}%
\bibitem [{\citenamefont {Dean}\ \emph {et~al.}(2013)\citenamefont {Dean},
  \citenamefont {Wang}, \citenamefont {Maher}, \citenamefont {Forsythe},
  \citenamefont {Ghahari}, \citenamefont {Gao}, \citenamefont {Katoch},
  \citenamefont {Ishigami}, \citenamefont {Moon}, \citenamefont {Koshino},
  \citenamefont {Taniguchi}, \citenamefont {Watanabe}, \citenamefont {Shepard},
  \citenamefont {Hone},\ and\ \citenamefont {Kim}}]{Dean2013_hof}%
  \BibitemOpen
  \bibfield  {author} {\bibinfo {author} {\bibfnamefont {C.~R.}\ \bibnamefont
  {Dean}}, \bibinfo {author} {\bibfnamefont {L.}~\bibnamefont {Wang}}, \bibinfo
  {author} {\bibfnamefont {P.}~\bibnamefont {Maher}}, \bibinfo {author}
  {\bibfnamefont {C.}~\bibnamefont {Forsythe}}, \bibinfo {author}
  {\bibfnamefont {F.}~\bibnamefont {Ghahari}}, \bibinfo {author} {\bibfnamefont
  {Y.}~\bibnamefont {Gao}}, \bibinfo {author} {\bibfnamefont {J.}~\bibnamefont
  {Katoch}}, \bibinfo {author} {\bibfnamefont {M.}~\bibnamefont {Ishigami}},
  \bibinfo {author} {\bibfnamefont {P.}~\bibnamefont {Moon}}, \bibinfo {author}
  {\bibfnamefont {M.}~\bibnamefont {Koshino}}, \bibinfo {author} {\bibfnamefont
  {T.}~\bibnamefont {Taniguchi}}, \bibinfo {author} {\bibfnamefont
  {K.}~\bibnamefont {Watanabe}}, \bibinfo {author} {\bibfnamefont {K.~L.}\
  \bibnamefont {Shepard}}, \bibinfo {author} {\bibfnamefont {J.}~\bibnamefont
  {Hone}}, \ and\ \bibinfo {author} {\bibfnamefont {P.}~\bibnamefont {Kim}},\
  }\bibfield  {title} {\enquote {\bibinfo {title} {Hofstadter's butterfly and
  the fractal quantum hall effect in moire superlattices},}\ }\href {\doibase
  10.1038/nature12186} {\bibfield  {journal} {\bibinfo  {journal} {Nature}\
  }\textbf {\bibinfo {volume} {497}},\ \bibinfo {pages} {598--602} (\bibinfo
  {year} {2013})}\BibitemShut {NoStop}%
\bibitem [{\citenamefont {Hunt}\ \emph {et~al.}(2013)\citenamefont {Hunt},
  \citenamefont {Sanchez-Yamagishi}, \citenamefont {Young}, \citenamefont
  {Yankowitz}, \citenamefont {LeRoy}, \citenamefont {Watanabe}, \citenamefont
  {Taniguchi}, \citenamefont {Moon}, \citenamefont {Koshino}, \citenamefont
  {Jarillo-Herrero},\ and\ \citenamefont {Ashoori}}]{Hunt2013_hof}%
  \BibitemOpen
  \bibfield  {author} {\bibinfo {author} {\bibfnamefont {B.}~\bibnamefont
  {Hunt}}, \bibinfo {author} {\bibfnamefont {J.~D.}\ \bibnamefont
  {Sanchez-Yamagishi}}, \bibinfo {author} {\bibfnamefont {A.~F.}\ \bibnamefont
  {Young}}, \bibinfo {author} {\bibfnamefont {M.}~\bibnamefont {Yankowitz}},
  \bibinfo {author} {\bibfnamefont {B.~J.}\ \bibnamefont {LeRoy}}, \bibinfo
  {author} {\bibfnamefont {K.}~\bibnamefont {Watanabe}}, \bibinfo {author}
  {\bibfnamefont {T.}~\bibnamefont {Taniguchi}}, \bibinfo {author}
  {\bibfnamefont {P.}~\bibnamefont {Moon}}, \bibinfo {author} {\bibfnamefont
  {M.}~\bibnamefont {Koshino}}, \bibinfo {author} {\bibfnamefont
  {P.}~\bibnamefont {Jarillo-Herrero}}, \ and\ \bibinfo {author} {\bibfnamefont
  {R.~C.}\ \bibnamefont {Ashoori}},\ }\bibfield  {title} {\enquote {\bibinfo
  {title} {Massive dirac fermions and hofstadter butterfly in a van der waals
  heterostructure},}\ }\href {\doibase 10.1126/science.1237240} {\bibfield
  {journal} {\bibinfo  {journal} {Science}\ }\textbf {\bibinfo {volume}
  {340}},\ \bibinfo {pages} {1427--1430} (\bibinfo {year} {2013})}\BibitemShut
  {NoStop}%
\bibitem [{\citenamefont {Castro~Neto}\ \emph {et~al.}(2009)\citenamefont
  {Castro~Neto}, \citenamefont {Guinea}, \citenamefont {Peres}, \citenamefont
  {Novoselov},\ and\ \citenamefont {Geim}}]{CastroNeto2009}%
  \BibitemOpen
  \bibfield  {author} {\bibinfo {author} {\bibfnamefont {A.~H.}\ \bibnamefont
  {Castro~Neto}}, \bibinfo {author} {\bibfnamefont {F.}~\bibnamefont {Guinea}},
  \bibinfo {author} {\bibfnamefont {N.~M.~R.}\ \bibnamefont {Peres}}, \bibinfo
  {author} {\bibfnamefont {K.~S.}\ \bibnamefont {Novoselov}}, \ and\ \bibinfo
  {author} {\bibfnamefont {A.~K.}\ \bibnamefont {Geim}},\ }\bibfield  {title}
  {\enquote {\bibinfo {title} {The electronic properties of graphene},}\ }\href
  {\doibase 10.1103/RevModPhys.81.109} {\bibfield  {journal} {\bibinfo
  {journal} {Rev. Mod. Phys.}\ }\textbf {\bibinfo {volume} {81}},\ \bibinfo
  {pages} {109} (\bibinfo {year} {2009})}\BibitemShut {NoStop}%
\bibitem [{\citenamefont {Chen}\ \emph
  {et~al.}(2008{\natexlab{a}})\citenamefont {Chen}, \citenamefont {Jang},
  \citenamefont {Adam}, \citenamefont {Fuhrer}, \citenamefont {Williams},\ and\
  \citenamefont {Ishigami}}]{Chen2008_cimp}%
  \BibitemOpen
  \bibfield  {author} {\bibinfo {author} {\bibfnamefont {J.~H.}\ \bibnamefont
  {Chen}}, \bibinfo {author} {\bibfnamefont {C.}~\bibnamefont {Jang}}, \bibinfo
  {author} {\bibfnamefont {S.}~\bibnamefont {Adam}}, \bibinfo {author}
  {\bibfnamefont {M.~S.}\ \bibnamefont {Fuhrer}}, \bibinfo {author}
  {\bibfnamefont {E.~D.}\ \bibnamefont {Williams}}, \ and\ \bibinfo {author}
  {\bibfnamefont {M.}~\bibnamefont {Ishigami}},\ }\bibfield  {title} {\enquote
  {\bibinfo {title} {Charged-impurity scattering in graphene},}\ }\href@noop {}
  {\bibfield  {journal} {\bibinfo  {journal} {Nat. Phys.}\ }\textbf {\bibinfo
  {volume} {4}},\ \bibinfo {pages} {377--381} (\bibinfo {year}
  {2008}{\natexlab{a}})}\BibitemShut {NoStop}%
\bibitem [{\citenamefont {Chen}\ \emph {et~al.}(2009)\citenamefont {Chen},
  \citenamefont {Cullen}, \citenamefont {Jang}, \citenamefont {Fuhrer},\ and\
  \citenamefont {Williams}}]{Chen2009_vacancy}%
  \BibitemOpen
  \bibfield  {author} {\bibinfo {author} {\bibfnamefont {J.-H.}\ \bibnamefont
  {Chen}}, \bibinfo {author} {\bibfnamefont {W.~G.}\ \bibnamefont {Cullen}},
  \bibinfo {author} {\bibfnamefont {C.}~\bibnamefont {Jang}}, \bibinfo {author}
  {\bibfnamefont {M.~S.}\ \bibnamefont {Fuhrer}}, \ and\ \bibinfo {author}
  {\bibfnamefont {E.~D.}\ \bibnamefont {Williams}},\ }\bibfield  {title}
  {\enquote {\bibinfo {title} {Defect scattering in graphene},}\ }\href
  {\doibase 10.1103/PhysRevLett.102.236805} {\bibfield  {journal} {\bibinfo
  {journal} {Phys. Rev. Lett.}\ }\textbf {\bibinfo {volume} {102}},\ \bibinfo
  {pages} {236805} (\bibinfo {year} {2009})}\BibitemShut {NoStop}%
\bibitem [{\citenamefont {Adam}\ \emph {et~al.}(2007)\citenamefont {Adam},
  \citenamefont {Hwang}, \citenamefont {Galitski},\ and\ \citenamefont
  {Das~Sarma}}]{Adam2007_Pnas}%
  \BibitemOpen
  \bibfield  {author} {\bibinfo {author} {\bibfnamefont {S.}~\bibnamefont
  {Adam}}, \bibinfo {author} {\bibfnamefont {E.~H.}\ \bibnamefont {Hwang}},
  \bibinfo {author} {\bibfnamefont {V.~M.}\ \bibnamefont {Galitski}}, \ and\
  \bibinfo {author} {\bibfnamefont {S.}~\bibnamefont {Das~Sarma}},\ }\bibfield
  {title} {\enquote {\bibinfo {title} {A self-consistent theory for graphene
  transport},}\ }\href {\doibase 10.1073/pnas.0704772104} {\bibfield  {journal}
  {\bibinfo  {journal} {Proc. Natl. Acad. Sci. USA}\ }\textbf {\bibinfo
  {volume} {104}},\ \bibinfo {pages} {18392} (\bibinfo {year}
  {2007})}\BibitemShut {NoStop}%
\bibitem [{\citenamefont {Jang}\ \emph {et~al.}(2008)\citenamefont {Jang},
  \citenamefont {Adam}, \citenamefont {Chen}, \citenamefont {Williams},
  \citenamefont {Das~Sarma},\ and\ \citenamefont {Fuhrer}}]{Jang2008}%
  \BibitemOpen
  \bibfield  {author} {\bibinfo {author} {\bibfnamefont {C.}~\bibnamefont
  {Jang}}, \bibinfo {author} {\bibfnamefont {S.}~\bibnamefont {Adam}}, \bibinfo
  {author} {\bibfnamefont {J.-H.}\ \bibnamefont {Chen}}, \bibinfo {author}
  {\bibfnamefont {E.~D.}\ \bibnamefont {Williams}}, \bibinfo {author}
  {\bibfnamefont {S.}~\bibnamefont {Das~Sarma}}, \ and\ \bibinfo {author}
  {\bibfnamefont {M.~S.}\ \bibnamefont {Fuhrer}},\ }\bibfield  {title}
  {\enquote {\bibinfo {title} {Tuning the effective fine structure constant in
  graphene: Opposing effects of dielectric screening on short- and long-range
  potential scattering},}\ }\href {\doibase 10.1103/PhysRevLett.101.146805}
  {\bibfield  {journal} {\bibinfo  {journal} {Phys. Rev. Lett.}\ }\textbf
  {\bibinfo {volume} {101}},\ \bibinfo {pages} {146805} (\bibinfo {year}
  {2008})}\BibitemShut {NoStop}%
\bibitem [{\citenamefont {Ponomarenko}\ \emph {et~al.}(2009)\citenamefont
  {Ponomarenko}, \citenamefont {Yang}, \citenamefont {Mohiuddin}, \citenamefont
  {Katsnelson}, \citenamefont {Novoselov}, \citenamefont {Morozov},
  \citenamefont {Zhukov}, \citenamefont {Schedin}, \citenamefont {Hill},\ and\
  \citenamefont {Geim}}]{Ponomarenko2009_highK}%
  \BibitemOpen
  \bibfield  {author} {\bibinfo {author} {\bibfnamefont {L.~A.}\ \bibnamefont
  {Ponomarenko}}, \bibinfo {author} {\bibfnamefont {R.}~\bibnamefont {Yang}},
  \bibinfo {author} {\bibfnamefont {T.~M.}\ \bibnamefont {Mohiuddin}}, \bibinfo
  {author} {\bibfnamefont {M.~I.}\ \bibnamefont {Katsnelson}}, \bibinfo
  {author} {\bibfnamefont {K.~S.}\ \bibnamefont {Novoselov}}, \bibinfo {author}
  {\bibfnamefont {S.~V.}\ \bibnamefont {Morozov}}, \bibinfo {author}
  {\bibfnamefont {A.~A.}\ \bibnamefont {Zhukov}}, \bibinfo {author}
  {\bibfnamefont {F.}~\bibnamefont {Schedin}}, \bibinfo {author} {\bibfnamefont
  {E.~W.}\ \bibnamefont {Hill}}, \ and\ \bibinfo {author} {\bibfnamefont
  {A.~K.}\ \bibnamefont {Geim}},\ }\bibfield  {title} {\enquote {\bibinfo
  {title} {Effect of a high-$\kappa$ environment on charge carrier mobility in
  graphene},}\ }\href {\doibase 10.1103/PhysRevLett.102.206603} {\bibfield
  {journal} {\bibinfo  {journal} {Phys. Rev. Lett.}\ }\textbf {\bibinfo
  {volume} {102}},\ \bibinfo {pages} {206603} (\bibinfo {year}
  {2009})}\BibitemShut {NoStop}%
\bibitem [{\citenamefont {Peres}(2010)}]{Peres2010_review}%
  \BibitemOpen
  \bibfield  {author} {\bibinfo {author} {\bibfnamefont {N.~M.~R.}\
  \bibnamefont {Peres}},\ }\bibfield  {title} {\enquote {\bibinfo {title}
  {Colloquium: The transport properties of graphene: An introduction},}\ }\href
  {\doibase 10.1103/RevModPhys.82.2673} {\bibfield  {journal} {\bibinfo
  {journal} {Rev. Mod. Phys.}\ }\textbf {\bibinfo {volume} {82}},\ \bibinfo
  {pages} {2673} (\bibinfo {year} {2010})}\BibitemShut {NoStop}%
\bibitem [{\citenamefont {Monteverde}\ \emph {et~al.}(2010)\citenamefont
  {Monteverde}, \citenamefont {Ojeda-Aristizabal}, \citenamefont {Weil},
  \citenamefont {Bennaceur}, \citenamefont {Ferrier}, \citenamefont {Gu\'eron},
  \citenamefont {Glattli}, \citenamefont {Bouchiat}, \citenamefont {Fuchs},\
  and\ \citenamefont {Maslov}}]{Monteverde2010_prl}%
  \BibitemOpen
  \bibfield  {author} {\bibinfo {author} {\bibfnamefont {M.}~\bibnamefont
  {Monteverde}}, \bibinfo {author} {\bibfnamefont {C.}~\bibnamefont
  {Ojeda-Aristizabal}}, \bibinfo {author} {\bibfnamefont {R.}~\bibnamefont
  {Weil}}, \bibinfo {author} {\bibfnamefont {K.}~\bibnamefont {Bennaceur}},
  \bibinfo {author} {\bibfnamefont {M.}~\bibnamefont {Ferrier}}, \bibinfo
  {author} {\bibfnamefont {S.}~\bibnamefont {Gu\'eron}}, \bibinfo {author}
  {\bibfnamefont {C.}~\bibnamefont {Glattli}}, \bibinfo {author} {\bibfnamefont
  {H.}~\bibnamefont {Bouchiat}}, \bibinfo {author} {\bibfnamefont {J.~N.}\
  \bibnamefont {Fuchs}}, \ and\ \bibinfo {author} {\bibfnamefont {D.~L.}\
  \bibnamefont {Maslov}},\ }\bibfield  {title} {\enquote {\bibinfo {title}
  {Transport and elastic scattering times as probes of the nature of impurity
  scattering in single-layer and bilayer graphene},}\ }\href {\doibase
  10.1103/PhysRevLett.104.126801} {\bibfield  {journal} {\bibinfo  {journal}
  {Phys. Rev. Lett.}\ }\textbf {\bibinfo {volume} {104}},\ \bibinfo {pages}
  {126801} (\bibinfo {year} {2010})}\BibitemShut {NoStop}%
\bibitem [{\citenamefont {Das~Sarma}\ \emph {et~al.}(2011)\citenamefont
  {Das~Sarma}, \citenamefont {Adam}, \citenamefont {Hwang},\ and\ \citenamefont
  {Rossi}}]{DasSarma2011_review}%
  \BibitemOpen
  \bibfield  {author} {\bibinfo {author} {\bibfnamefont {S.}~\bibnamefont
  {Das~Sarma}}, \bibinfo {author} {\bibfnamefont {S.}~\bibnamefont {Adam}},
  \bibinfo {author} {\bibfnamefont {E.~H.}\ \bibnamefont {Hwang}}, \ and\
  \bibinfo {author} {\bibfnamefont {E.}~\bibnamefont {Rossi}},\ }\bibfield
  {title} {\enquote {\bibinfo {title} {Electronic transport in two-dimensional
  graphene},}\ }\href {\doibase 10.1103/RevModPhys.83.407} {\bibfield
  {journal} {\bibinfo  {journal} {Rev. Mod. Phys.}\ }\textbf {\bibinfo {volume}
  {83}},\ \bibinfo {pages} {407} (\bibinfo {year} {2011})}\BibitemShut
  {NoStop}%
\bibitem [{\citenamefont {Couto}\ \emph {et~al.}(2011)\citenamefont {Couto},
  \citenamefont {Sacepe},\ and\ \citenamefont {Morpurgo}}]{Couto2011}%
  \BibitemOpen
  \bibfield  {author} {\bibinfo {author} {\bibfnamefont {N.~J.~G.}\
  \bibnamefont {Couto}}, \bibinfo {author} {\bibfnamefont {B.}~\bibnamefont
  {Sacepe}}, \ and\ \bibinfo {author} {\bibfnamefont {A.~F.}\ \bibnamefont
  {Morpurgo}},\ }\bibfield  {title} {\enquote {\bibinfo {title} {Transport
  through graphene on srtio$_3$},}\ }\href@noop {} {\bibfield  {journal}
  {\bibinfo  {journal} {Phys. Rev. Lett.}\ }\textbf {\bibinfo {volume} {107}},\
  \bibinfo {pages} {225501} (\bibinfo {year} {2011})}\BibitemShut {NoStop}%
\bibitem [{\citenamefont {Adam}\ \emph {et~al.}(2011)\citenamefont {Adam},
  \citenamefont {Jung}, \citenamefont {Klimov}, \citenamefont {Zhitenev},
  \citenamefont {Stroscio},\ and\ \citenamefont {Stiles}}]{Aetal11}%
  \BibitemOpen
  \bibfield  {author} {\bibinfo {author} {\bibfnamefont {S.}~\bibnamefont
  {Adam}}, \bibinfo {author} {\bibfnamefont {S.}~\bibnamefont {Jung}}, \bibinfo
  {author} {\bibfnamefont {N.~N.}\ \bibnamefont {Klimov}}, \bibinfo {author}
  {\bibfnamefont {N.~B.}\ \bibnamefont {Zhitenev}}, \bibinfo {author}
  {\bibfnamefont {J.~A.}\ \bibnamefont {Stroscio}}, \ and\ \bibinfo {author}
  {\bibfnamefont {M.~D.}\ \bibnamefont {Stiles}},\ }\bibfield  {title}
  {\enquote {\bibinfo {title} {Mechanism for puddle formation in graphene},}\
  }\href {\doibase 10.1103/PhysRevB.84.235421} {\bibfield  {journal} {\bibinfo
  {journal} {Phys. Rev. B}\ }\textbf {\bibinfo {volume} {84}},\ \bibinfo
  {pages} {235421} (\bibinfo {year} {2011})}\BibitemShut {NoStop}%
\bibitem [{\citenamefont {Neumann}\ \emph {et~al.}(2014)\citenamefont
  {Neumann}, \citenamefont {Reichardt}, \citenamefont {Drogeler}, \citenamefont
  {Watanabe}, \citenamefont {Taniguchi}, \citenamefont {Beschoten},
  \citenamefont {Rotkin},\ and\ \citenamefont {Stampfer}}]{neu14}%
  \BibitemOpen
  \bibfield  {author} {\bibinfo {author} {\bibfnamefont {C.}~\bibnamefont
  {Neumann}}, \bibinfo {author} {\bibfnamefont {S.}~\bibnamefont {Reichardt}},
  \bibinfo {author} {\bibfnamefont {M.}~\bibnamefont {Drogeler}}, \bibinfo
  {author} {\bibfnamefont {K.}~\bibnamefont {Watanabe}}, \bibinfo {author}
  {\bibfnamefont {T.}~\bibnamefont {Taniguchi}}, \bibinfo {author}
  {\bibfnamefont {B.}~\bibnamefont {Beschoten}}, \bibinfo {author}
  {\bibfnamefont {S.~V.}\ \bibnamefont {Rotkin}}, \ and\ \bibinfo {author}
  {\bibfnamefont {C.}~\bibnamefont {Stampfer}},\ }\bibfield  {title} {\enquote
  {\bibinfo {title} {Magneto-raman microscopy for probing local material
  properties of graphene},}\ }\href@noop {} {\bibfield  {journal} {\bibinfo
  {journal} {ArXiv e-prints}\ } (\bibinfo {year} {2014})},\ \Eprint
  {http://arxiv.org/abs/1406.7771} {arXiv:1406.7771 [cond-mat.mes-hall]}
  \BibitemShut {NoStop}%
\bibitem [{\citenamefont {Taychatanapat}\ \emph {et~al.}(2013)\citenamefont
  {Taychatanapat}, \citenamefont {Watanabe}, \citenamefont {Taniguchi},\ and\
  \citenamefont {Jarillo-Herrero}}]{Taychatanapat2013_Bfocus}%
  \BibitemOpen
  \bibfield  {author} {\bibinfo {author} {\bibfnamefont {T.}~\bibnamefont
  {Taychatanapat}}, \bibinfo {author} {\bibfnamefont {K.}~\bibnamefont
  {Watanabe}}, \bibinfo {author} {\bibfnamefont {T.}~\bibnamefont {Taniguchi}},
  \ and\ \bibinfo {author} {\bibfnamefont {P.}~\bibnamefont
  {Jarillo-Herrero}},\ }\bibfield  {title} {\enquote {\bibinfo {title}
  {Electrically tunable transverse magnetic focusing in graphene},}\
  }\href@noop {} {\bibfield  {journal} {\bibinfo  {journal} {Nat. Phys.}\
  }\textbf {\bibinfo {volume} {9}},\ \bibinfo {pages} {225--229} (\bibinfo
  {year} {2013})}\BibitemShut {NoStop}%
\bibitem [{\citenamefont {Chen}\ \emph
  {et~al.}(2008{\natexlab{b}})\citenamefont {Chen}, \citenamefont {Jang},
  \citenamefont {Xiao}, \citenamefont {Ishigami},\ and\ \citenamefont
  {Fuhrer}}]{Chen2008}%
  \BibitemOpen
  \bibfield  {author} {\bibinfo {author} {\bibfnamefont {J.-H.}\ \bibnamefont
  {Chen}}, \bibinfo {author} {\bibfnamefont {C.}~\bibnamefont {Jang}}, \bibinfo
  {author} {\bibfnamefont {S.}~\bibnamefont {Xiao}}, \bibinfo {author}
  {\bibfnamefont {M.}~\bibnamefont {Ishigami}}, \ and\ \bibinfo {author}
  {\bibfnamefont {M.~S.}\ \bibnamefont {Fuhrer}},\ }\bibfield  {title}
  {\enquote {\bibinfo {title} {Intrinsic and extrinsic performance limits of
  graphene devices on sio$_2$},}\ }\href@noop {} {\bibfield  {journal}
  {\bibinfo  {journal} {Nat. Nano.}\ }\textbf {\bibinfo {volume} {3}},\
  \bibinfo {pages} {206--209} (\bibinfo {year}
  {2008}{\natexlab{b}})}\BibitemShut {NoStop}%
\bibitem [{\citenamefont {Martin}\ \emph {et~al.}(2008)\citenamefont {Martin},
  \citenamefont {Akerman}, \citenamefont {Ulbricht}, \citenamefont {Lohmann},
  \citenamefont {Smet}, \citenamefont {von Klitzing},\ and\ \citenamefont
  {Yacoby}}]{Martin2008}%
  \BibitemOpen
  \bibfield  {author} {\bibinfo {author} {\bibfnamefont {J.}~\bibnamefont
  {Martin}}, \bibinfo {author} {\bibfnamefont {N.}~\bibnamefont {Akerman}},
  \bibinfo {author} {\bibfnamefont {G.}~\bibnamefont {Ulbricht}}, \bibinfo
  {author} {\bibfnamefont {T.}~\bibnamefont {Lohmann}}, \bibinfo {author}
  {\bibfnamefont {J.~H.}\ \bibnamefont {Smet}}, \bibinfo {author}
  {\bibfnamefont {K.}~\bibnamefont {von Klitzing}}, \ and\ \bibinfo {author}
  {\bibfnamefont {A.}~\bibnamefont {Yacoby}},\ }\bibfield  {title} {\enquote
  {\bibinfo {title} {Observation of electron-hole puddles in graphene using a
  scanning single-electron transistor},}\ }\href@noop {} {\bibfield  {journal}
  {\bibinfo  {journal} {Nat. Phys.}\ }\textbf {\bibinfo {volume} {4}},\
  \bibinfo {pages} {144--148} (\bibinfo {year} {2008})}\BibitemShut {NoStop}%
\bibitem [{Note1()}]{Note1}%
  \BibitemOpen
  \bibinfo {note} {In Ref.~\cite {Couto2011} we discussed transport through
  graphene-on-SrTiO$_3$ in the context of resonant scattering, but we also
  pointed out --in Ref.~(21) of that paper-- that the data are compatible with
  scattering by ripples, i.e. with the conclusions of this present
  work}\BibitemShut {NoStop}%
\bibitem [{\citenamefont {McCann}\ \emph {et~al.}(2006)\citenamefont {McCann},
  \citenamefont {Kechedzhi}, \citenamefont {Fal'ko}, \citenamefont {Suzuura},
  \citenamefont {Ando},\ and\ \citenamefont {Altshuler}}]{McCann2006}%
  \BibitemOpen
  \bibfield  {author} {\bibinfo {author} {\bibfnamefont {E.}~\bibnamefont
  {McCann}}, \bibinfo {author} {\bibfnamefont {K.}~\bibnamefont {Kechedzhi}},
  \bibinfo {author} {\bibfnamefont {Vladimir~I.}\ \bibnamefont {Fal'ko}},
  \bibinfo {author} {\bibfnamefont {H.}~\bibnamefont {Suzuura}}, \bibinfo
  {author} {\bibfnamefont {T.}~\bibnamefont {Ando}}, \ and\ \bibinfo {author}
  {\bibfnamefont {B.~L.}\ \bibnamefont {Altshuler}},\ }\bibfield  {title}
  {\enquote {\bibinfo {title} {Weak-localization magnetoresistance and valley
  symmetry in graphene},}\ }\href {\doibase 10.1103/PhysRevLett.97.146805}
  {\bibfield  {journal} {\bibinfo  {journal} {Phys. Rev. Lett.}\ }\textbf
  {\bibinfo {volume} {97}},\ \bibinfo {pages} {146805} (\bibinfo {year}
  {2006})}\BibitemShut {NoStop}%
\bibitem [{\citenamefont {Morpurgo}\ and\ \citenamefont
  {Guinea}(2006)}]{Morpurgo2006}%
  \BibitemOpen
  \bibfield  {author} {\bibinfo {author} {\bibfnamefont {A.~F.}\ \bibnamefont
  {Morpurgo}}\ and\ \bibinfo {author} {\bibfnamefont {F.}~\bibnamefont
  {Guinea}},\ }\bibfield  {title} {\enquote {\bibinfo {title} {Intervalley
  scattering, long-range disorder, and effective time-reversal symmetry
  breaking in graphene},}\ }\href@noop {} {\bibfield  {journal} {\bibinfo
  {journal} {Phys. Rev. Lett.}\ }\textbf {\bibinfo {volume} {97}},\ \bibinfo
  {pages} {196804} (\bibinfo {year} {2006})}\BibitemShut {NoStop}%
\bibitem [{\citenamefont {Tikhonenko}\ \emph {et~al.}(2008)\citenamefont
  {Tikhonenko}, \citenamefont {Horsell}, \citenamefont {Gorbachev},\ and\
  \citenamefont {Savchenko}}]{Tikhonenko2008}%
  \BibitemOpen
  \bibfield  {author} {\bibinfo {author} {\bibfnamefont {F.~V.}\ \bibnamefont
  {Tikhonenko}}, \bibinfo {author} {\bibfnamefont {D.~W.}\ \bibnamefont
  {Horsell}}, \bibinfo {author} {\bibfnamefont {R.~V.}\ \bibnamefont
  {Gorbachev}}, \ and\ \bibinfo {author} {\bibfnamefont {A.~K.}\ \bibnamefont
  {Savchenko}},\ }\bibfield  {title} {\enquote {\bibinfo {title} {Weak
  localization in graphene flakes},}\ }\href@noop {} {\bibfield  {journal}
  {\bibinfo  {journal} {Phys. Rev. Lett.}\ }\textbf {\bibinfo {volume} {100}},\
  \bibinfo {pages} {056802} (\bibinfo {year} {2008})}\BibitemShut {NoStop}%
\bibitem [{\citenamefont {Guignard}\ \emph {et~al.}(2012)\citenamefont
  {Guignard}, \citenamefont {Leprat}, \citenamefont {Glattli}, \citenamefont
  {Schopfer},\ and\ \citenamefont {Poirier}}]{Guignard2012}%
  \BibitemOpen
  \bibfield  {author} {\bibinfo {author} {\bibfnamefont {J.}~\bibnamefont
  {Guignard}}, \bibinfo {author} {\bibfnamefont {D.}~\bibnamefont {Leprat}},
  \bibinfo {author} {\bibfnamefont {D.~C.}\ \bibnamefont {Glattli}}, \bibinfo
  {author} {\bibfnamefont {F.}~\bibnamefont {Schopfer}}, \ and\ \bibinfo
  {author} {\bibfnamefont {W.}~\bibnamefont {Poirier}},\ }\bibfield  {title}
  {\enquote {\bibinfo {title} {Quantum hall effect in exfoliated graphene
  affected by charged impurities: Metrological measurements},}\ }\href@noop {}
  {\bibfield  {journal} {\bibinfo  {journal} {Phys. Rev. B}\ }\textbf {\bibinfo
  {volume} {85}},\ \bibinfo {pages} {165420--} (\bibinfo {year}
  {2012})}\BibitemShut {NoStop}%
\bibitem [{\citenamefont {Ferry}\ and\ \citenamefont
  {Goodnick}(2009)}]{FerryBook2009}%
  \BibitemOpen
  \bibfield  {author} {\bibinfo {author} {\bibfnamefont {D.~K.}\ \bibnamefont
  {Ferry}}\ and\ \bibinfo {author} {\bibfnamefont {S.~M.}\ \bibnamefont
  {Goodnick}},\ }\href@noop {} {\emph {\bibinfo {title} {Transport in
  Nanostructures}}}\ (\bibinfo  {publisher} {Cambridge University Press},\
  \bibinfo {year} {2009})\BibitemShut {NoStop}%
\bibitem [{\citenamefont {Baker}\ \emph {et~al.}(2012)\citenamefont {Baker},
  \citenamefont {Alexander-Webber}, \citenamefont {Altebaeumer}, \citenamefont
  {Janssen}, \citenamefont {Tzalenchuk}, \citenamefont {Lara-Avila},
  \citenamefont {Kubatkin}, \citenamefont {Yakimova}, \citenamefont {Lin},
  \citenamefont {Li},\ and\ \citenamefont {Nicholas}}]{Baker2012}%
  \BibitemOpen
  \bibfield  {author} {\bibinfo {author} {\bibfnamefont {A.~M.~R.}\
  \bibnamefont {Baker}}, \bibinfo {author} {\bibfnamefont {J.~A.}\ \bibnamefont
  {Alexander-Webber}}, \bibinfo {author} {\bibfnamefont {T.}~\bibnamefont
  {Altebaeumer}}, \bibinfo {author} {\bibfnamefont {T.~J. B.~M.}\ \bibnamefont
  {Janssen}}, \bibinfo {author} {\bibfnamefont {A.}~\bibnamefont {Tzalenchuk}},
  \bibinfo {author} {\bibfnamefont {S.}~\bibnamefont {Lara-Avila}}, \bibinfo
  {author} {\bibfnamefont {S.}~\bibnamefont {Kubatkin}}, \bibinfo {author}
  {\bibfnamefont {R.}~\bibnamefont {Yakimova}}, \bibinfo {author}
  {\bibfnamefont {C.-T.}\ \bibnamefont {Lin}}, \bibinfo {author} {\bibfnamefont
  {L.-J.}\ \bibnamefont {Li}}, \ and\ \bibinfo {author} {\bibfnamefont {R.~J.}\
  \bibnamefont {Nicholas}},\ }\bibfield  {title} {\enquote {\bibinfo {title}
  {Weak localization scattering lengths in epitaxial, and cvd graphene},}\
  }\href@noop {} {\bibfield  {journal} {\bibinfo  {journal} {Phys. Rev. B}\
  }\textbf {\bibinfo {volume} {86}},\ \bibinfo {pages} {235441} (\bibinfo
  {year} {2012})}\BibitemShut {NoStop}%
\bibitem [{\citenamefont {{Engels}}\ \emph {et~al.}(2014)\citenamefont
  {{Engels}}, \citenamefont {{Terr{\'e}s}}, \citenamefont {{Epping}},
  \citenamefont {{Khodkov}}, \citenamefont {{Watanabe}}, \citenamefont
  {{Taniguchi}}, \citenamefont {{Beschoten}},\ and\ \citenamefont
  {{Stampfer}}}]{engels14}%
  \BibitemOpen
  \bibfield  {author} {\bibinfo {author} {\bibfnamefont {S.}~\bibnamefont
  {{Engels}}}, \bibinfo {author} {\bibfnamefont {B.}~\bibnamefont
  {{Terr{\'e}s}}}, \bibinfo {author} {\bibfnamefont {A.}~\bibnamefont
  {{Epping}}}, \bibinfo {author} {\bibfnamefont {T.}~\bibnamefont {{Khodkov}}},
  \bibinfo {author} {\bibfnamefont {K.}~\bibnamefont {{Watanabe}}}, \bibinfo
  {author} {\bibfnamefont {T.}~\bibnamefont {{Taniguchi}}}, \bibinfo {author}
  {\bibfnamefont {B.}~\bibnamefont {{Beschoten}}}, \ and\ \bibinfo {author}
  {\bibfnamefont {C.}~\bibnamefont {{Stampfer}}},\ }\bibfield  {title}
  {\enquote {\bibinfo {title} {{Limitations to carrier mobility and
  phase-coherent transport in bilayer graphene}},}\ }\href@noop {} {\bibfield
  {journal} {\bibinfo  {journal} {ArXiv e-prints}\ } (\bibinfo {year}
  {2014})},\ \Eprint {http://arxiv.org/abs/1403.1547} {arXiv:1403.1547
  [cond-mat.mes-hall]} \BibitemShut {NoStop}%
\bibitem [{\citenamefont {Ando}(2006)}]{Ando}%
  \BibitemOpen
  \bibfield  {author} {\bibinfo {author} {\bibfnamefont {T.}~\bibnamefont
  {Ando}},\ }\bibfield  {title} {\enquote {\bibinfo {title} {Screening effect
  and impurity scattering in monolayer graphene},}\ }\href@noop {} {\bibfield
  {journal} {\bibinfo  {journal} {J. Phys. Soc. Jpn.}\ }\textbf {\bibinfo
  {volume} {75}},\ \bibinfo {pages} {074716} (\bibinfo {year}
  {2006})}\BibitemShut {NoStop}%
\bibitem [{\citenamefont {Nomura}\ and\ \citenamefont
  {MacDonald}(2007)}]{Nomura2007}%
  \BibitemOpen
  \bibfield  {author} {\bibinfo {author} {\bibfnamefont {K.}~\bibnamefont
  {Nomura}}\ and\ \bibinfo {author} {\bibfnamefont {A.~H.}\ \bibnamefont
  {MacDonald}},\ }\bibfield  {title} {\enquote {\bibinfo {title} {Quantum
  transport of massless dirac fermions},}\ }\href@noop {} {\bibfield  {journal}
  {\bibinfo  {journal} {Phys. Rev. Lett.}\ }\textbf {\bibinfo {volume} {98}},\
  \bibinfo {pages} {076602} (\bibinfo {year} {2007})}\BibitemShut {NoStop}%
\bibitem [{\citenamefont {Katsnelson}\ and\ \citenamefont
  {Geim}(2008)}]{katsnelson_corrug}%
  \BibitemOpen
  \bibfield  {author} {\bibinfo {author} {\bibfnamefont {M.~I.}\ \bibnamefont
  {Katsnelson}}\ and\ \bibinfo {author} {\bibfnamefont {A.~K.}\ \bibnamefont
  {Geim}},\ }\bibfield  {title} {\enquote {\bibinfo {title} {Electron
  scattering on microscopic corrugations in graphene},}\ }\href@noop {}
  {\bibfield  {journal} {\bibinfo  {journal} {Phil. Trans. R. Soc. A}\ }\textbf
  {\bibinfo {volume} {366}},\ \bibinfo {pages} {195--204} (\bibinfo {year}
  {2008})}\BibitemShut {NoStop}%
\bibitem [{\citenamefont {Vozmediano}\ \emph {et~al.}(2010)\citenamefont
  {Vozmediano}, \citenamefont {Katsnelson},\ and\ \citenamefont
  {Guinea}}]{Vozmediano2010}%
  \BibitemOpen
  \bibfield  {author} {\bibinfo {author} {\bibfnamefont {M.~A.~H.}\
  \bibnamefont {Vozmediano}}, \bibinfo {author} {\bibfnamefont {M.~I.}\
  \bibnamefont {Katsnelson}}, \ and\ \bibinfo {author} {\bibfnamefont
  {F.}~\bibnamefont {Guinea}},\ }\bibfield  {title} {\enquote {\bibinfo {title}
  {Gauge fields in graphene},}\ }\href@noop {} {\bibfield  {journal} {\bibinfo
  {journal} {Phys. Rep.}\ }\textbf {\bibinfo {volume} {496}},\ \bibinfo {pages}
  {109--148} (\bibinfo {year} {2010})}\BibitemShut {NoStop}%
\bibitem [{\citenamefont {Graf}\ \emph {et~al.}(2007)\citenamefont {Graf},
  \citenamefont {Molitor}, \citenamefont {Ensslin}, \citenamefont {Stampfer},
  \citenamefont {Jungen}, \citenamefont {Hierold},\ and\ \citenamefont
  {Wirtz}}]{graf07}%
  \BibitemOpen
  \bibfield  {author} {\bibinfo {author} {\bibfnamefont {D.}~\bibnamefont
  {Graf}}, \bibinfo {author} {\bibfnamefont {F.}~\bibnamefont {Molitor}},
  \bibinfo {author} {\bibfnamefont {K.}~\bibnamefont {Ensslin}}, \bibinfo
  {author} {\bibfnamefont {C.}~\bibnamefont {Stampfer}}, \bibinfo {author}
  {\bibfnamefont {A.}~\bibnamefont {Jungen}}, \bibinfo {author} {\bibfnamefont
  {C.}~\bibnamefont {Hierold}}, \ and\ \bibinfo {author} {\bibfnamefont
  {L.}~\bibnamefont {Wirtz}},\ }\bibfield  {title} {\enquote {\bibinfo {title}
  {Spatially resolved raman spectroscopy of single- and few-layer graphene},}\
  }\href {\doibase 10.1021/nl061702a} {\bibfield  {journal} {\bibinfo
  {journal} {Nano Lett.}\ }\textbf {\bibinfo {volume} {7}},\ \bibinfo {pages}
  {238--242} (\bibinfo {year} {2007})}\BibitemShut {NoStop}%
\bibitem [{\citenamefont {Lee}\ \emph {et~al.}(2012)\citenamefont {Lee},
  \citenamefont {Ahn}, \citenamefont {Shim}, \citenamefont {Lee},\ and\
  \citenamefont {Ryu}}]{lee12}%
  \BibitemOpen
  \bibfield  {author} {\bibinfo {author} {\bibfnamefont {J.~E.}\ \bibnamefont
  {Lee}}, \bibinfo {author} {\bibfnamefont {G.}~\bibnamefont {Ahn}}, \bibinfo
  {author} {\bibfnamefont {J.}~\bibnamefont {Shim}}, \bibinfo {author}
  {\bibfnamefont {Y.~S.}\ \bibnamefont {Lee}}, \ and\ \bibinfo {author}
  {\bibfnamefont {S.}~\bibnamefont {Ryu}},\ }\bibfield  {title} {\enquote
  {\bibinfo {title} {Optical separation of mechanical strain from charge doping
  in graphene},}\ }\href@noop {} {\bibfield  {journal} {\bibinfo  {journal}
  {Nat. Commun.}\ }\textbf {\bibinfo {volume} {3}},\ \bibinfo {pages} {1024}
  (\bibinfo {year} {2012})}\BibitemShut {NoStop}%
\bibitem [{\citenamefont {Pisana}\ \emph {et~al.}(2007)\citenamefont {Pisana},
  \citenamefont {Lazzeri}, \citenamefont {Casiraghi}, \citenamefont
  {Novoselov}, \citenamefont {Geim}, \citenamefont {Ferrari},\ and\
  \citenamefont {Mauri}}]{pisana07}%
  \BibitemOpen
  \bibfield  {author} {\bibinfo {author} {\bibfnamefont {S.}~\bibnamefont
  {Pisana}}, \bibinfo {author} {\bibfnamefont {M.}~\bibnamefont {Lazzeri}},
  \bibinfo {author} {\bibfnamefont {C.}~\bibnamefont {Casiraghi}}, \bibinfo
  {author} {\bibfnamefont {K.~S.}\ \bibnamefont {Novoselov}}, \bibinfo {author}
  {\bibfnamefont {A.~K.}\ \bibnamefont {Geim}}, \bibinfo {author}
  {\bibfnamefont {A.~C.}\ \bibnamefont {Ferrari}}, \ and\ \bibinfo {author}
  {\bibfnamefont {F.}~\bibnamefont {Mauri}},\ }\bibfield  {title} {\enquote
  {\bibinfo {title} {Breakdown of the adiabatic born-oppenheimer approximation
  in graphene},}\ }\href@noop {} {\bibfield  {journal} {\bibinfo  {journal}
  {Nat. Mater.}\ }\textbf {\bibinfo {volume} {6}},\ \bibinfo {pages} {198--201}
  (\bibinfo {year} {2007})}\BibitemShut {NoStop}%
\bibitem [{\citenamefont {Stampfer}\ \emph {et~al.}(2007)\citenamefont
  {Stampfer}, \citenamefont {Molitor}, \citenamefont {Graf}, \citenamefont
  {Ensslin}, \citenamefont {Jungen}, \citenamefont {Hierold},\ and\
  \citenamefont {Wirtz}}]{stampfer07}%
  \BibitemOpen
  \bibfield  {author} {\bibinfo {author} {\bibfnamefont {C.}~\bibnamefont
  {Stampfer}}, \bibinfo {author} {\bibfnamefont {F.}~\bibnamefont {Molitor}},
  \bibinfo {author} {\bibfnamefont {D.}~\bibnamefont {Graf}}, \bibinfo {author}
  {\bibfnamefont {K.}~\bibnamefont {Ensslin}}, \bibinfo {author} {\bibfnamefont
  {A.}~\bibnamefont {Jungen}}, \bibinfo {author} {\bibfnamefont
  {C.}~\bibnamefont {Hierold}}, \ and\ \bibinfo {author} {\bibfnamefont
  {L.}~\bibnamefont {Wirtz}},\ }\bibfield  {title} {\enquote {\bibinfo {title}
  {Raman imaging of doping domains in graphene on sio2},}\ }\href@noop {}
  {\bibfield  {journal} {\bibinfo  {journal} {Appl. Phys. Lett.}\ }\textbf
  {\bibinfo {volume} {91}},\ \bibinfo {pages} {241907} (\bibinfo {year}
  {2007})}\BibitemShut {NoStop}%
\bibitem [{\citenamefont {Berciaud}\ \emph {et~al.}(2013)\citenamefont
  {Berciaud}, \citenamefont {Li}, \citenamefont {Htoon}, \citenamefont {Brus},
  \citenamefont {Doorn},\ and\ \citenamefont {Heinz}}]{berciaud13}%
  \BibitemOpen
  \bibfield  {author} {\bibinfo {author} {\bibfnamefont {S.}~\bibnamefont
  {Berciaud}}, \bibinfo {author} {\bibfnamefont {X.}~\bibnamefont {Li}},
  \bibinfo {author} {\bibfnamefont {H.}~\bibnamefont {Htoon}}, \bibinfo
  {author} {\bibfnamefont {L.~E.}\ \bibnamefont {Brus}}, \bibinfo {author}
  {\bibfnamefont {S.~K.}\ \bibnamefont {Doorn}}, \ and\ \bibinfo {author}
  {\bibfnamefont {T.~F.}\ \bibnamefont {Heinz}},\ }\bibfield  {title} {\enquote
  {\bibinfo {title} {Intrinsic line shape of the raman 2d-mode in freestanding
  graphene monolayers},}\ }\href {\doibase 10.1021/nl400917e} {\bibfield
  {journal} {\bibinfo  {journal} {Nano Lett.}\ }\textbf {\bibinfo {volume}
  {13}},\ \bibinfo {pages} {3517--3523} (\bibinfo {year} {2013})}\BibitemShut
  {NoStop}%
\bibitem [{\citenamefont {Popov}\ and\ \citenamefont {Lambin}(2013)}]{popov13}%
  \BibitemOpen
  \bibfield  {author} {\bibinfo {author} {\bibfnamefont {V.~N.}\ \bibnamefont
  {Popov}}\ and\ \bibinfo {author} {\bibfnamefont {P.}~\bibnamefont {Lambin}},\
  }\bibfield  {title} {\enquote {\bibinfo {title} {Theoretical 2d raman band of
  strained graphene},}\ }\href {\doibase 10.1103/PhysRevB.87.155425} {\bibfield
   {journal} {\bibinfo  {journal} {Phys. Rev. B}\ }\textbf {\bibinfo {volume}
  {87}},\ \bibinfo {pages} {155425} (\bibinfo {year} {2013})}\BibitemShut
  {NoStop}%
\bibitem [{\citenamefont {Forster}\ \emph {et~al.}(2013)\citenamefont
  {Forster}, \citenamefont {Molina-Sanchez}, \citenamefont {Engels},
  \citenamefont {Epping}, \citenamefont {Watanabe}, \citenamefont {Taniguchi},
  \citenamefont {Wirtz},\ and\ \citenamefont {Stampfer}}]{forster13}%
  \BibitemOpen
  \bibfield  {author} {\bibinfo {author} {\bibfnamefont {F.}~\bibnamefont
  {Forster}}, \bibinfo {author} {\bibfnamefont {A.}~\bibnamefont
  {Molina-Sanchez}}, \bibinfo {author} {\bibfnamefont {S.}~\bibnamefont
  {Engels}}, \bibinfo {author} {\bibfnamefont {A.}~\bibnamefont {Epping}},
  \bibinfo {author} {\bibfnamefont {K.}~\bibnamefont {Watanabe}}, \bibinfo
  {author} {\bibfnamefont {T.}~\bibnamefont {Taniguchi}}, \bibinfo {author}
  {\bibfnamefont {L.}~\bibnamefont {Wirtz}}, \ and\ \bibinfo {author}
  {\bibfnamefont {C.}~\bibnamefont {Stampfer}},\ }\bibfield  {title} {\enquote
  {\bibinfo {title} {Dielectric screening of the kohn anomaly of graphene on
  hexagonal boron nitride},}\ }\href {\doibase 10.1103/PhysRevB.88.085419}
  {\bibfield  {journal} {\bibinfo  {journal} {Phys. Rev. B}\ }\textbf {\bibinfo
  {volume} {88}},\ \bibinfo {pages} {085419} (\bibinfo {year}
  {2013})}\BibitemShut {NoStop}%
\bibitem [{\citenamefont {Zakharchenko}\ \emph {et~al.}(2009)\citenamefont
  {Zakharchenko}, \citenamefont {Katsnelson},\ and\ \citenamefont
  {Fasolino}}]{Zakharchenko2009}%
  \BibitemOpen
  \bibfield  {author} {\bibinfo {author} {\bibfnamefont {K.~V.}\ \bibnamefont
  {Zakharchenko}}, \bibinfo {author} {\bibfnamefont {M.~I.}\ \bibnamefont
  {Katsnelson}}, \ and\ \bibinfo {author} {\bibfnamefont {A.}~\bibnamefont
  {Fasolino}},\ }\bibfield  {title} {\enquote {\bibinfo {title} {Finite
  temperature lattice properties of graphene beyond the quasiharmonic
  approximation},}\ }\href@noop {} {\bibfield  {journal} {\bibinfo  {journal}
  {Phys. Rev. Lett.}\ }\textbf {\bibinfo {volume} {102}},\ \bibinfo {pages}
  {046808--} (\bibinfo {year} {2009})}\BibitemShut {NoStop}%
\bibitem [{\citenamefont {Levy}\ \emph {et~al.}(2010)\citenamefont {Levy},
  \citenamefont {Burke}, \citenamefont {Meaker}, \citenamefont {Panlasigui},
  \citenamefont {Zettl}, \citenamefont {Guinea}, \citenamefont {Castro~Neto},\
  and\ \citenamefont {Crommie}}]{Letal1o}%
  \BibitemOpen
  \bibfield  {author} {\bibinfo {author} {\bibfnamefont {N.}~\bibnamefont
  {Levy}}, \bibinfo {author} {\bibfnamefont {S.~A.}\ \bibnamefont {Burke}},
  \bibinfo {author} {\bibfnamefont {K.~L.}\ \bibnamefont {Meaker}}, \bibinfo
  {author} {\bibfnamefont {M.}~\bibnamefont {Panlasigui}}, \bibinfo {author}
  {\bibfnamefont {A.}~\bibnamefont {Zettl}}, \bibinfo {author} {\bibfnamefont
  {F.}~\bibnamefont {Guinea}}, \bibinfo {author} {\bibfnamefont {A.~H.}\
  \bibnamefont {Castro~Neto}}, \ and\ \bibinfo {author} {\bibfnamefont {M.~F.}\
  \bibnamefont {Crommie}},\ }\bibfield  {title} {\enquote {\bibinfo {title}
  {Strain-induced pseudo-magnetic fields greater than 300 tesla in graphene
  nanobubbles},}\ }\href {\doibase 10.1126/science.1191700} {\bibfield
  {journal} {\bibinfo  {journal} {Science}\ }\textbf {\bibinfo {volume}
  {329}},\ \bibinfo {pages} {544--547} (\bibinfo {year} {2010})}\BibitemShut
  {NoStop}%
\bibitem [{\citenamefont {Gibertini}\ \emph {et~al.}(2012)\citenamefont
  {Gibertini}, \citenamefont {Tomadin}, \citenamefont {Guinea}, \citenamefont
  {Katsnelson},\ and\ \citenamefont {Polini}}]{GTGKP12}%
  \BibitemOpen
  \bibfield  {author} {\bibinfo {author} {\bibfnamefont {M.}~\bibnamefont
  {Gibertini}}, \bibinfo {author} {\bibfnamefont {A.}~\bibnamefont {Tomadin}},
  \bibinfo {author} {\bibfnamefont {F.}~\bibnamefont {Guinea}}, \bibinfo
  {author} {\bibfnamefont {M.~I.}\ \bibnamefont {Katsnelson}}, \ and\ \bibinfo
  {author} {\bibfnamefont {M.}~\bibnamefont {Polini}},\ }\bibfield  {title}
  {\enquote {\bibinfo {title} {Electron-hole puddles in the absence of charged
  impurities},}\ }\href {\doibase 10.1103/PhysRevB.85.201405} {\bibfield
  {journal} {\bibinfo  {journal} {Phys. Rev. B}\ }\textbf {\bibinfo {volume}
  {85}},\ \bibinfo {pages} {201405} (\bibinfo {year} {2012})}\BibitemShut
  {NoStop}%
\bibitem [{\citenamefont {Ono}\ and\ \citenamefont {Sugihara}(1966)}]{OS66}%
  \BibitemOpen
  \bibfield  {author} {\bibinfo {author} {\bibfnamefont {S.}~\bibnamefont
  {Ono}}\ and\ \bibinfo {author} {\bibfnamefont {K.}~\bibnamefont {Sugihara}},\
  }\bibfield  {title} {\enquote {\bibinfo {title} {Theory of the transport
  properties in graphite},}\ }\href {\doibase 10.1143/JPSJ.21.861} {\bibfield
  {journal} {\bibinfo  {journal} {J. Phys. Soc. Jpn.}\ }\textbf {\bibinfo
  {volume} {21}},\ \bibinfo {pages} {861--868} (\bibinfo {year}
  {1966})}\BibitemShut {NoStop}%
\bibitem [{\citenamefont {Suzuura}\ and\ \citenamefont {Ando}(2002)}]{SA02b}%
  \BibitemOpen
  \bibfield  {author} {\bibinfo {author} {\bibfnamefont {H.}~\bibnamefont
  {Suzuura}}\ and\ \bibinfo {author} {\bibfnamefont {T.}~\bibnamefont {Ando}},\
  }\bibfield  {title} {\enquote {\bibinfo {title} {Phonons and electron-phonon
  scattering in carbon nanotubes},}\ }\href {\doibase
  10.1103/PhysRevB.65.235412} {\bibfield  {journal} {\bibinfo  {journal} {Phys.
  Rev. B}\ }\textbf {\bibinfo {volume} {65}},\ \bibinfo {pages} {235412}
  (\bibinfo {year} {2002})}\BibitemShut {NoStop}%
\bibitem [{\citenamefont {Choi}\ \emph {et~al.}(2010)\citenamefont {Choi},
  \citenamefont {Jhi},\ and\ \citenamefont {Son}}]{CJS10}%
  \BibitemOpen
  \bibfield  {author} {\bibinfo {author} {\bibfnamefont {S.-M.}\ \bibnamefont
  {Choi}}, \bibinfo {author} {\bibfnamefont {S.-H.}\ \bibnamefont {Jhi}}, \
  and\ \bibinfo {author} {\bibfnamefont {Y.-W.}\ \bibnamefont {Son}},\
  }\bibfield  {title} {\enquote {\bibinfo {title} {Effects of strain on
  electronic properties of graphene},}\ }\href {\doibase
  10.1103/PhysRevB.81.081407} {\bibfield  {journal} {\bibinfo  {journal} {Phys.
  Rev. B}\ }\textbf {\bibinfo {volume} {81}},\ \bibinfo {pages} {081407}
  (\bibinfo {year} {2010})}\BibitemShut {NoStop}%
\bibitem [{\citenamefont {Woods}\ \emph {et~al.}(2014)\citenamefont {Woods},
  \citenamefont {Britnell}, \citenamefont {Eckmann}, \citenamefont {Ma},
  \citenamefont {Lu}, \citenamefont {Guo}, \citenamefont {Lin}, \citenamefont
  {Yu}, \citenamefont {Cao}, \citenamefont {Gorbachev}, \citenamefont
  {Kretinin}, \citenamefont {Park}, \citenamefont {Ponomarenko}, \citenamefont
  {Katsnelson}, \citenamefont {Gornostyrev}, \citenamefont {Watanabe},
  \citenamefont {Taniguchi}, \citenamefont {Casiraghi}, \citenamefont {Gao},
  \citenamefont {Geim},\ and\ \citenamefont {Novoselov}}]{Wetal14o}%
  \BibitemOpen
  \bibfield  {author} {\bibinfo {author} {\bibfnamefont {C.~R.}\ \bibnamefont
  {Woods}}, \bibinfo {author} {\bibfnamefont {L.}~\bibnamefont {Britnell}},
  \bibinfo {author} {\bibfnamefont {A.}~\bibnamefont {Eckmann}}, \bibinfo
  {author} {\bibfnamefont {R.~S.}\ \bibnamefont {Ma}}, \bibinfo {author}
  {\bibfnamefont {J.~C.}\ \bibnamefont {Lu}}, \bibinfo {author} {\bibfnamefont
  {H.~M.}\ \bibnamefont {Guo}}, \bibinfo {author} {\bibfnamefont
  {X.}~\bibnamefont {Lin}}, \bibinfo {author} {\bibfnamefont {G.~L.}\
  \bibnamefont {Yu}}, \bibinfo {author} {\bibfnamefont {Y.}~\bibnamefont
  {Cao}}, \bibinfo {author} {\bibfnamefont {R.~V.}\ \bibnamefont {Gorbachev}},
  \bibinfo {author} {\bibfnamefont {A.~V.}\ \bibnamefont {Kretinin}}, \bibinfo
  {author} {\bibfnamefont {J.}~\bibnamefont {Park}}, \bibinfo {author}
  {\bibfnamefont {L.~A.}\ \bibnamefont {Ponomarenko}}, \bibinfo {author}
  {\bibfnamefont {M.~I.}\ \bibnamefont {Katsnelson}}, \bibinfo {author}
  {\bibfnamefont {Yu~N.}\ \bibnamefont {Gornostyrev}}, \bibinfo {author}
  {\bibfnamefont {K.}~\bibnamefont {Watanabe}}, \bibinfo {author}
  {\bibfnamefont {T.}~\bibnamefont {Taniguchi}}, \bibinfo {author}
  {\bibfnamefont {C.}~\bibnamefont {Casiraghi}}, \bibinfo {author}
  {\bibfnamefont {H.~J.}\ \bibnamefont {Gao}}, \bibinfo {author} {\bibfnamefont
  {A.~K.}\ \bibnamefont {Geim}}, \ and\ \bibinfo {author} {\bibfnamefont
  {K.~S.}\ \bibnamefont {Novoselov}},\ }\bibfield  {title} {\enquote {\bibinfo
  {title} {Commensurate-incommensurate transition in graphene on hexagonal
  boron nitride},}\ }\href {\doibase 10.1038/nphys2954} {\bibfield  {journal}
  {\bibinfo  {journal} {Nat. Phys.}\ }\textbf {\bibinfo {volume} {10}},\
  \bibinfo {pages} {451--456} (\bibinfo {year} {2014})}\BibitemShut {NoStop}%
\bibitem [{\citenamefont {{Jung}}\ \emph {et~al.}(2014)\citenamefont {{Jung}},
  \citenamefont {{DaSilva}}, \citenamefont {{Adam}},\ and\ \citenamefont
  {{MacDonald}}}]{JDAM14}%
  \BibitemOpen
  \bibfield  {author} {\bibinfo {author} {\bibfnamefont {J.}~\bibnamefont
  {{Jung}}}, \bibinfo {author} {\bibfnamefont {A.}~\bibnamefont {{DaSilva}}},
  \bibinfo {author} {\bibfnamefont {S.}~\bibnamefont {{Adam}}}, \ and\ \bibinfo
  {author} {\bibfnamefont {A.~H.}\ \bibnamefont {{MacDonald}}},\ }\bibfield
  {title} {\enquote {\bibinfo {title} {Origin of band gaps in graphene on
  hexagonal boron nitride},}\ }\href@noop {} {\bibfield  {journal} {\bibinfo
  {journal} {ArXiv e-prints}\ } (\bibinfo {year} {2014})},\ \Eprint
  {http://arxiv.org/abs/1403.0496} {arXiv:1403.0496 [cond-mat.mes-hall]}
  \BibitemShut {NoStop}%
\bibitem [{\citenamefont {San-Jose}\ \emph {et~al.}(2014)\citenamefont
  {San-Jose}, \citenamefont {Guti\'errez-Rubio}, \citenamefont {Sturla},\ and\
  \citenamefont {Guinea}}]{SGSG14}%
  \BibitemOpen
  \bibfield  {author} {\bibinfo {author} {\bibfnamefont {P.}~\bibnamefont
  {San-Jose}}, \bibinfo {author} {\bibfnamefont {A.}~\bibnamefont
  {Guti\'errez-Rubio}}, \bibinfo {author} {\bibfnamefont {M.}~\bibnamefont
  {Sturla}}, \ and\ \bibinfo {author} {\bibfnamefont {F.}~\bibnamefont
  {Guinea}},\ }\bibfield  {title} {\enquote {\bibinfo {title} {Spontaneous
  strains and gap in graphene on boron nitride},}\ }\href {\doibase
  10.1103/PhysRevB.90.075428} {\bibfield  {journal} {\bibinfo  {journal} {Phys.
  Rev. B}\ }\textbf {\bibinfo {volume} {90}},\ \bibinfo {pages} {075428}
  (\bibinfo {year} {2014})}\BibitemShut {NoStop}%
\end{thebibliography}

%

\end{document}